\documentclass[lineno]{JFM-FLM_Au}
\usepackage{subfig}
\usepackage{amsmath}
\usepackage{graphicx}
\usepackage{float}
\usepackage{hyperref}
\usepackage{cleveref}

\lefttitle{Analytical model for stall cell formation}
\righttitle{Journal of Fluid Mechanics}

\title{Stall cells over an airfoil. Part 2: A vortex-based analytical model for their formation and saturation}

\author{Rishabh Mishra\aff{1}, Emmanuel Guilmineau\aff{1}, Ingrid Neunaber\aff{2}, and Caroline Braud \aff{1}}

\affiliation{\aff{1}LHEEA lab. - CNRS - Nantes Université, Centrale Nantes, 1 rue de la No\"e,
44100 Nantes, France
\aff{2} FLOW, Department of Engineering Mechanics, KTH Royal Institute of Technology, SE-100 44 Stockholm, Sweden}

\corresau{Emmanuel Guilmineau, \email{emmanuel.guilmineau@ec-nantes.fr}}

\begin{document}
\nolinenumbers
\maketitle

\begin{abstract}
Stall cells are spanwise-periodic flow structures that spontaneously form on airfoils operating near stall, fundamentally altering the aerodynamic loading distribution. Despite decades of experimental observations, a complete theoretical framework connecting vortex dynamics to the characteristic flow patterns has remained elusive. In this work, we develop an analytical model for stall cell formation based on the interaction between finite-length, counter-rotating vortex tubes representing the separation vortex and trailing-edge vortex. Linear stability analysis of the coupled vortex system yields the growth rate and wavelength selection of the Crow-type instability responsible for the wave-like bending of the vortex structures. A weakly nonlinear analysis using the method of multiple scales is performed to derive the Stuart--Landau amplitude equation, providing an explicit expression for the saturation amplitude at which nonlinear effects arrest the instability growth and establish quasi-steady cellular structures. The vortex sheet representing the separated shear layer is coupled to the vortex tube dynamics through the Birkhoff--Rott equation, from which we derive the induced vertical vorticity $\Omega_y$ that drives the alternating spanwise velocity characteristic of stall cells. The model predicts quantitatively the spanwise velocity magnitude, vertical vorticity distribution, and vortex sheet deformation. The resulting framework provides a unified, first-principles description connecting the Crow-type instability of counter-rotating vortex tubes to the observed flow topology of stall cells. The model is validated against the DDES simulation data presented in the companion paper \cite{PartI}, demonstrating strong agreement.
\end{abstract}

\newpage

\section{Introduction}
\label{sec:introduction}

The aerodynamic stall of lifting surfaces remains one of the most consequential phenomena in fluid mechanics, with profound implications for aircraft safety, wind turbine performance, and a wide range of engineering applications involving separated flows. Stall is classically understood as the sudden loss of lift experienced by an airfoil when the angle of attack exceeds a critical value, triggering massive flow separation on the suction side \citep{McCroskey1982}. While this two-dimensional description captures the essential features of the lift breakdown, the actual stall process on finite wings and high-aspect-ratio lifting surfaces is inherently three-dimensional, exhibiting complex spatiotemporal dynamics that have challenged researchers for over five decades \citep{Zaman1989,Broeren2001}.

Among the most striking manifestations of three-dimensionality in stalled flows is the spontaneous formation of \emph{stall cells}---periodic, cellular patterns that appear on the suction side of airfoils operating near stall conditions. These structures, first systematically documented by \citet{Gregory1971} and \citet{Moss1971}, are characterised by spanwise-periodic regions of attached and separated flow that produce distinctive surface flow topologies often described as ``owl-faced'' or ``mushroom-shaped'' patterns \citep{Winkelmann1982}. The presence of stall cells fundamentally alters the aerodynamic loading distribution, introduces low-frequency unsteadiness, and can significantly impact the performance and structural integrity of lifting surfaces \citep{Disotell2016,Broeren2001}.

Despite their practical importance and the substantial body of experimental evidence accumulated over the years, a complete theoretical understanding of stall cell formation has remained elusive. Several competing hypotheses have been advanced to explain the physical mechanisms underlying this phenomenon, yet no unified framework exists that connects the vortex dynamics of the separated shear layer to the characteristic spanwise flow patterns observed in experiments and simulations. The present work aims to address this gap by developing an analytical model based on the interaction between finite-length vortex tubes, extending classical vortex stability theory to derive quantitative predictions for the onset, wavelength selection, and saturation amplitude of stall cells.

\subsection{Experimental observations and phenomenology}

The experimental study of stall cells has a rich history spanning more than fifty years. The pioneering observations of \citet{Gregory1971} revealed three-dimensional flow patterns during stall development on aerofoils tested in wind tunnels, where the wingspan occupied the entire test section width. Initially, these cellular structures were attributed to potential artifacts arising from sidewall boundary layer interactions. This hypothesis was definitively refuted by the subsequent work of \citet{Winkelman1980} and \citet{Winkelmann1982}, who demonstrated that stall cells form on finite aspect-ratio wings with free tips, establishing them as intrinsic features of the separated flow rather than facility-dependent phenomena.

The morphological characteristics of stall cells have been extensively documented through surface flow visualization, pressure measurements, and velocity field surveys. \citet{Winkelman1980} proposed a ``tentative flowfield model'' describing the time-averaged topology, wherein counter-rotating vortex pairs emanate from focal points on the airfoil surface and extend into the wake. This model was refined by subsequent investigations revealing that stall cell vortices originate approximately normal to the wing surface and gradually align with the freestream direction as they convect downstream \citep{Manolesos2014a,Manolesos2014b}. The characteristic spanwise wavelength of stall cells typically ranges from one to two chord lengths, though considerable variation has been reported depending on the airfoil profile, Reynolds number, and angle of attack \citep{Schewe2001,Manolesos2013}.

The parametric dependence of stall cell formation has been systematically investigated by several research groups. \citet{Schewe2001} demonstrated significant Reynolds number effects on the flow patterns around bluff bodies and airfoils near stall. \citet{Manolesos2013} established that the critical angle of attack for stall cell onset depends primarily on the airfoil profile rather than the aspect ratio, and identified a relationship between the critical angle and Reynolds number. More recently, \citet{DellOrso2018} performed a comprehensive parametric study identifying eight distinct flow topologies and documented conditions under which the flow exhibits bistability, oscillating between different topological states.

The unsteady dynamics of stall cells have attracted considerable attention due to their potential impact on structural loads and aeroacoustic emissions. \citet{Yon1998} observed that stall cells undergo ``jostling'' motions and trail downstream in an unsteady manner. \citet{Disotell2015} and \citet{Disotell2016} employed pressure-sensitive paint to detect large-amplitude pressure fluctuations associated with cellular separation, revealing low-frequency oscillations that couple with the global aerodynamic forces. \cite{hanna2026trackingstallcelldynamics} recently showed that these low-frequency oscillations exist around the stall cell borders, a phenomenon that cannot be captured from global load measurements.
These large-amplitude fluctuations were recently observed by \cite{Neunaber2022}, and \cite{BraudPhysreview2024} to be spatially highly anti-correlated in the spanwise direction, leading to wall pressure bi-stability responsible for large load fluctuations. This was later linked to the unsteady displacement of the stall cell in the spanwise direction (see \cite{hanna2026trackingstallcelldynamics}). Their dynamics at Reynolds numbers larger than $10^6$ are dominated by a coherent motion in the spanwise direction with a characteristic velocity of order $0.1U$. The motion can be decomposed into a large-scale, low-frequency sweep with a Strouhal number $St \sim 0.001$, combined with faster, smaller-scale oscillations. 

\subsection{Theoretical models: the Crow instability hypothesis}

The first theoretical attempt to explain stall cell formation from a vortex dynamics perspective was made by \citet{Weihs1983}, who proposed that the Crow instability, a classical long-wavelength instability of counter-rotating vortex pairs, may be responsible for the emergence of cellular patterns in poststall flows. The Crow instability, originally discovered by \citet{Crow1970} in the context of aircraft wake vortices, describes the mutual amplification of sinusoidal perturbations on a pair of parallel, counter-rotating line vortices. The instability arises from the interaction between the strain field induced by one vortex on the other, leading to symmetric bending modes that grow exponentially until the vortices eventually link and form a series of vortex rings.

In the seminal analysis of \citet{Crow1970}, the vortices are idealized as interacting line singularities, with finite core effects incorporated through a cutoff in the self-induction integral. The resulting eigenvalue problem yields a most amplified wavelength of approximately $8.6b$, where $b$ is the vortex separation distance, with growth rates proportional to $\Gamma/b^2$, where $\Gamma$ is the circulation. The instability is most pronounced for long-wavelength, symmetric modes in which both vortices bend in phase within planes inclined at approximately $48^\circ$ to the horizontal. This theoretical framework has been extensively validated through laboratory experiments \citep{Widnall1971} and numerical simulations \citep{Han2000} of aircraft wake vortices.

\citet{Weihs1983} adapted this framework to the stall cell problem by considering the interaction between the separation vortex shed from the suction side and a counter-rotating vortex associated with the trailing edge flow. They derived a first-order estimate relating the wing aspect ratio to the number of cellular patterns, finding reasonable agreement with experimental observations. However, the \citet{Weihs1983} model suffers from several fundamental limitations. First, the analysis is restricted to the linear regime and provides no mechanism for amplitude saturation, leaving the finite-amplitude structure of stall cells unexplained. Second, the model does not explicitly account for the vortex sheet representing the separated shear layer, which plays a crucial role in the flow topology (see the companion paper \cite{PartI}). Third, the connection between vortex bending and the characteristic spanwise velocity patterns of stall cells is not established. As noted by \citet{Rodriguez2011}, the inviscid nature of the Crow instability does not directly predict the formation of stall cells, since flow separation is fundamentally a viscous phenomenon.

\subsection{Theoretical models: lifting-line theory approaches}

An alternative theoretical framework for understanding stall cells emerged from classical lifting-line theory. \citet{Spalart2014} developed a simple yet insightful model demonstrating that the uniform spanwise flow solution becomes unstable when the lift curve $C_L(\alpha)$ exhibits a negative slope. The physical reasoning is as follows: in the attached flow regime where $\partial C_L/\partial \alpha > 0$, spanwise perturbations in circulation are self-correcting through the induced velocity field. However, when the lift curve slope becomes negative (as occurs in the poststall regime), these perturbations are amplified, leading to the spontaneous formation of ``lift cells''---spanwise-periodic variations in the circulation distribution that correspond to the observed stall cells.

The \citet{Spalart2014} analysis successfully captures the qualitative features of stall cell onset and predicts spanwise lift distributions with nearly square-wave patterns, consistent with the sharp boundaries between attached and separated regions observed experimentally. The criterion $\partial C_L/\partial \alpha < 0$ has been widely adopted as a necessary condition for stall cell formation \citep{Plante2021,Shen2017}. Building on this framework, \citet{Gross2015} derived a formula for the spanwise wavelength of stall cells based on the balance between destabilizing and restoring effects associated with local circulation perturbations.

\citet{Gallay2015} examined nonlinear lifting-line algorithms for pre- and poststall flows, demonstrating that both circulation-based and angle-of-attack-based correction methods can reproduce multiple solutions in the poststall regime corresponding to different stall cell configurations. \citet{Plante2022} extended the inviscid analysis to a lifting-surface framework using the vortex lattice method. These lifting-line and lifting-surface approaches offer computational efficiency and physical insight, but they are inherently limited by the assumptions of potential flow theory and the empirical treatment of viscous effects through sectional lift curves. They do not resolve the vortex dynamics within the separated region and cannot capture the detailed flow topology associated with stall cells.

\subsection{Theoretical models: global stability analysis}

A fundamentally different approach to understanding stall cell formation has emerged from global linear stability analysis of the Navier--Stokes equations. \citet{Rodriguez2010,Rodriguez2011} performed pioneering analyses of the laminar flow around a Joukowski airfoil at low Reynolds numbers ($Re = 200$), identifying a three-dimensional stationary global eigenmode whose spatial structure, when superimposed on the two-dimensional base flow, produces surface streamlines strongly reminiscent of experimentally observed stall cells. Their work employed critical point theory to demonstrate that two-dimensional flow topologies are structurally unstable, and any three-dimensional perturbation, regardless of amplitude, leads to a qualitatively different flow topology.

The global stability framework provides a rigorous mathematical foundation for understanding the onset of three-dimensionality in separated flows. However, subsequent investigations revealed important complexities. \citet{He2017} performed comprehensive analyses of the NACA 0009, 0015, and 4415 airfoils at Reynolds numbers up to 1000, demonstrating that the leading modal instability is the two-dimensional B\'enard--von K\'arm\'an wake mode rather than the three-dimensional stationary mode associated with stall cells. Similar conclusions were reached by \citet{Zhang2016} for the NACA 0012 airfoil. These findings indicate that at low Reynolds numbers, stall-cell-like patterns emerge as secondary features rather than primary flow bifurcations, raising questions about the direct applicability of laminar stability results to the turbulent flows of practical interest.

To address stall cells at realistic Reynolds numbers ($Re \sim 10^5$), \citet{Plante2021} performed global stability analysis using the Reynolds-averaged Navier--Stokes (RANS) equations with the Spalart--Allmaras turbulence model \citep{Spalart1992}. Investigating the NACA 4412 airfoil at $Re = 350\,000$, they demonstrated that a three-dimensional stationary mode becomes the primary instability at $\alpha \approx 15^\circ$, and its nonlinear evolution leads to stall cell formation. More recently, \citet{Sarras2024} developed a data-consistent RANS approach using adjoint-based data assimilation to calibrate turbulence model corrections, substantially improving the agreement between predicted and observed critical angles.

While global stability analysis provides valuable insights into the modal structure of separated flows, this framework focuses primarily on the linear onset of instability and does not directly address the finite-amplitude saturated state that corresponds to fully developed stall cells. The connection between the linear eigenmodes and the observed flow topology remains largely phenomenological, relying on visualization of superposed perturbations rather than derivation from first principles.

\subsection{Motivating observations from high-fidelity simulations}

To guide the development of a theoretical model that addresses the limitations of existing approaches, we performed delayed detached-eddy simulations (DDES) of the flow over an airfoil derived from a wind turbine blade section at $Re_c = 2.0 \times 10^5$ and $14^\circ$ angle of attack. The computational domain spanned four chord lengths in the spanwise direction with an aspect ratio of 4. Details of the numerical methodology and comprehensive flow analysis are reported in a companion paper \cite{}; here we summarise the key observations that motivate the present theoretical development.

\subsubsection{Counter-rotating vortex tube pair and Crow-type bending}

\begin{figure}
\centering
\includegraphics[width=0.8\textwidth]{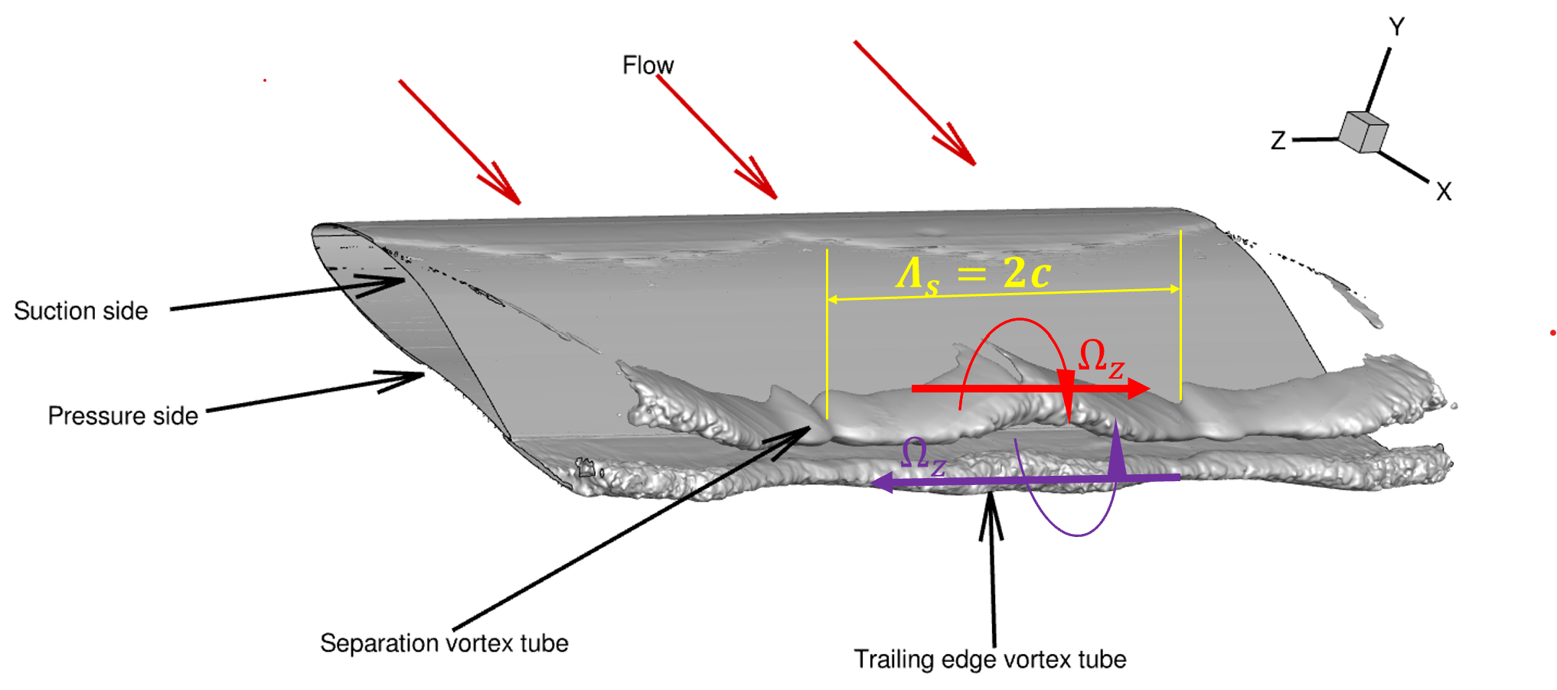}
\caption{Time-averaged $Q = 7$ iso-surface from DDES showing the separation vortex tube and trailing-edge vortex tube forming a counter-rotating pair. The wave-like bending of both structures is evident.}
\label{fig:vortex_tubes}
\end{figure}

The DDES results reveal that the separated shear layer rolls up into a coherent separation vortex tube characterised by strong spanwise vorticity $\Omega_z$. This separation vortex forms a counter-rotating pair with the trailing-edge vortex tube, as illustrated in figure~\ref{fig:vortex_tubes}. The time-averaged $Q$-criterion iso-surface ($Q = 7$) clearly shows both vortex tubes, with the separation vortex located above the trailing-edge vortex.

Crucially, both vortex tubes exhibit a pronounced wave-like bending in the spanwise direction, with wavelength $\Lambda_s \approx 2c$ equal to the stall cell wavelength. This bending is consistent with the Crow instability mechanism proposed by \citet{Weihs1983}, providing direct evidence that the mutual induction between the counter-rotating vortex pair drives the three-dimensional organization of the flow. The bending of the vortex tubes is correlated with a corresponding deformation of the vortex sheet (separated shear layer), which appears as a ``hump'' in the $Q$-criterion iso-surface near $z/c = 0$. For more details on this, look at the companion paper \cite{PartI}.

\subsubsection{Generation of vertical vorticity and spanwise velocity}

\begin{figure}
    \centering
    \includegraphics[width=\linewidth]{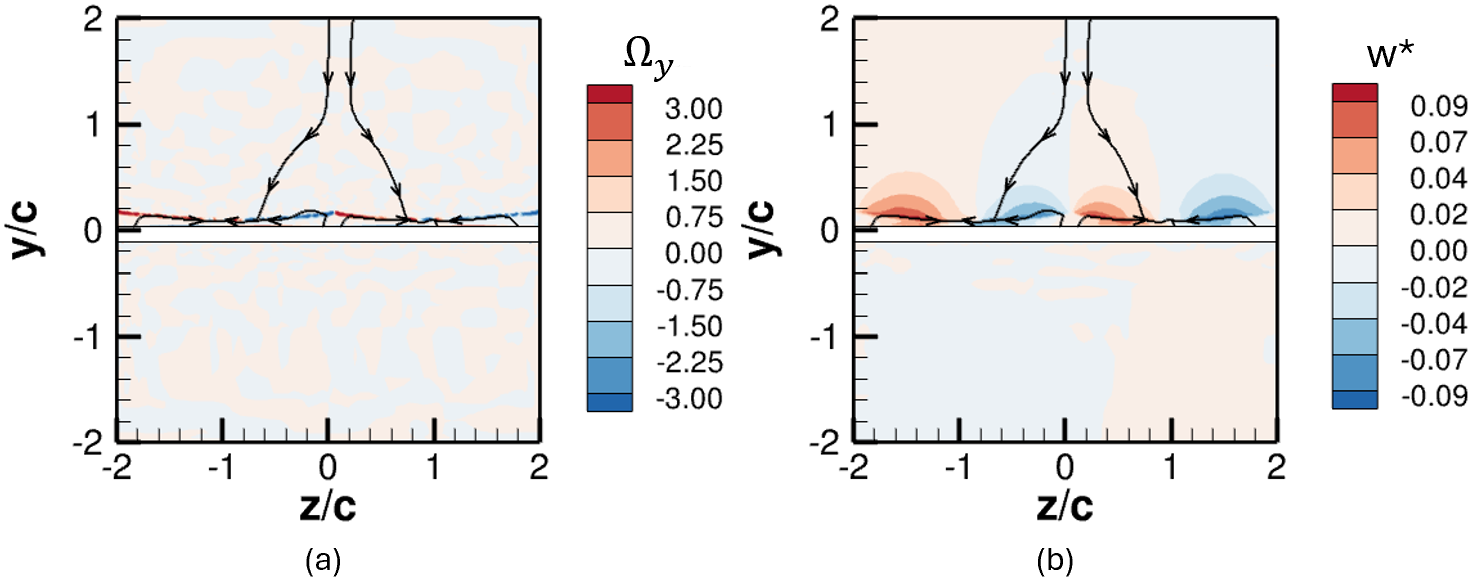}
    \caption{Contour plots from DDES at $x/c = 0.6$: (a) normalised vertical vorticity $\Omega_y$ showing alternating positive and negative regions; (b) normalised spanwise velocity $w^*$ exhibiting the characteristic stall cell pattern. The black curves are showing in-plane streamlines.}
    \label{fig:omega_y_w}
\end{figure}

A central finding of the DDES analysis is that the wave-like bending of the vortex tube and vortex sheet induces significant vertical vorticity $\Omega_y$. Figure~\ref{fig:omega_y_w} shows contour plots of $\Omega_y$ and the normalised spanwise velocity $w^* = W/U_\infty$ at $x/c = 0.6$. The $\Omega_y$ distribution exhibits alternating positive and negative regions in the spanwise direction, with the highest magnitudes occurring within the separated shear layer. This vertical vorticity component, which is absent in two-dimensional analyses, is directly responsible for inducing the spanwise velocity that characterises stall cells. For more details, refer to the companion paper \cite{PartI}.

The spanwise velocity field shows the characteristic stall cell pattern: alternating regions of positive and negative $w^*$ with magnitudes reaching $|w^*| \approx 0.5$. The stall cell cores, defined as regions of maximum $|w^*|$, are located near $z/c \approx 0$ and $z/c \approx \pm 1$, yielding a stall cell wavelength of $\Lambda_c \approx 2c$ consistent with experimental observations on thick airfoils \citep{Schewe2001,Yon1998}.





\subsubsection{Quasi-steady saturated state}

The DDES results demonstrate that the stall cells reach a quasi-steady saturated state at $14^\circ$ angle of attack, with the spanwise velocity structures maintaining their spatial organization over time. This observation indicates that nonlinear effects arrest the growth of the Crow-type instability at a finite amplitude, establishing a stable cellular pattern. In contrast, at higher angles of attack ($16^\circ$), the stall cells exhibit temporal instability with alternating dominance between different spanwise regions, suggesting proximity to a secondary bifurcation. For more information on this, look at the companion paper \cite{PartI}.

\subsection{Gaps in existing theoretical understanding}

The theoretical models reviewed above have each contributed important insights into stall cell formation, yet significant gaps remain in our understanding of this phenomenon. Based on the DDES observations, we identify the following key limitations of existing approaches:

\begin{enumerate}
\item \textbf{Lack of a saturation mechanism.} The Crow instability analysis of \citet{Weihs1983} and the lifting-line model of \citet{Spalart2014} predict the onset of instability but provide no mechanism for amplitude saturation. The DDES results show that stall cells reach a finite amplitude and persist as quasi-steady structures rather than growing without bound. The nonlinear processes that arrest the instability growth and establish the saturated state have not been theoretically characterised.

\item \textbf{Absence of explicit vortex sheet dynamics.} The separated shear layer, which rolls up to form the separation vortex, is represented in existing models either implicitly (through circulation distributions in lifting-line theory) or not at all (in the original Crow instability framework). The DDES results demonstrate that the coupling between vortex tube bending and shear layer deformation is central to the generation of $\Omega_y$ and the spanwise velocity patterns.

\item \textbf{No derivation of the vertical vorticity field.} The DDES results show that $\Omega_y$ is the key vorticity component driving the spanwise flow. No theoretical model has derived the $\Omega_y$ distribution from the underlying vortex dynamics or explained why its magnitude remains approximately constant with downstream distance.

\item \textbf{No derivation of the spanwise velocity field.} The characteristic signature of stall cells is an alternating spanwise velocity component. While this velocity pattern has been extensively documented experimentally and computationally, no theoretical model has derived its structure, magnitude, or spatial distribution from first principles.

\end{enumerate}

\subsection{Objectives and outline of the present work}

The present work aims to address these gaps by developing a comprehensive analytical model for stall cell formation based on the interaction between finite-length vortex tubes. Our approach extends the classical Crow instability framework in several important ways:

\begin{enumerate}
\item We consider \emph{finite-length} vortex filaments with sinusoidal perturbations satisfying pinned boundary conditions, capturing the spanwise confinement inherent to finite wings and computational domains.

\item We derive the full \emph{interaction matrix} governing the coupled dynamics of the separation vortex tube and trailing-edge vortex tube, obtaining analytical expressions for the growth rate of the instability.

\item We perform a \emph{weakly nonlinear analysis} using the method of multiple scales to derive the Stuart--Landau equation governing the amplitude evolution. This analysis yields explicit expressions for the Landau coefficient and the \emph{saturation amplitude} at which nonlinear effects arrest the instability growth.

\item We explicitly model the \emph{vortex sheet} representing the separated shear layer, coupling its deformation to the vortex tube bending through an attachment condition at the separation line. The sheet dynamics are governed by the Birkhoff--Rott equation, from which we derive the induced \emph{vertical vorticity} resulting from sheet tilting.

\item We compute the \emph{alternating spanwise velocity field} induced by the $\Omega_y$ distribution, demonstrating that the characteristic stall cell flow pattern emerges naturally from the vortex dynamics.

\item We derive analytical expressions for all \emph{saturated flow quantities}, sheet deformation, vertical vorticity, and spanwise velocity, and validate these predictions against DDES data.
\end{enumerate}

The resulting model provides a unified theoretical framework that connects the Crow-like instability of counter-rotating vortex tubes to the observed flow topology of stall cells, while yielding quantitative predictions that can be tested against experiments and simulations.

The remainder of this paper is organised as follows. Section~\ref{sec:model} presents the mathematical formulationof the model, beginning with the base configuration and governing equations, proceeding through the induced velocity calculations, perturbation analysis for the Crow-like instability, and weakly nonlinear theory for amplitude saturation. 
Section~\ref{sec:validation} presents validation of the model against DDES results, comparing predictions for the vortex sheet deformation, spanwise velocity, vertical vorticity, and the rotation angle at multiple streamwise locations. Conclusions are presented in section \ref{sec:conclusions}.

\section{Mathematical model of vortex tube interaction leading to stall cell formation} \label{sec:model}

In this section, we present a detailed derivation of a mathematical model describing the interaction between two finite-length vortex tubes in a simplified representation of the flow over an airfoil at high angles of attack, culminating in the formation of stall cells. The model elucidates the mechanism proposed herein: the mutual induction between a separation vortex tube originating from the suction side and a trailing edge vortex tube from the pressure side initiates a Crow-like instability, resulting in the bending of the separation vortex tube. This bending, in turn, deforms an attached vortex sheet representing the separation shear layer, thereby inducing alternating vertical vorticity (the \(y\)-component) along the spanwise direction (\(z\)). The alternating \(y\)-vorticity generates an oscillatory spanwise velocity field, which manifests as the characteristic signature of stall cells, periodic regions of separated flow along the span.

\noindent The derivation is grounded in the principles of inviscid, incompressible fluid dynamics, governed by the Euler equations, where vorticity is concentrated along the vortex tubes and the sheet. We opt for finite-length vortex tubes to more realistically capture end effects and spanwise confinement, unlike infinite-tube models that overlook boundary influences but simplify analytics; this choice allows for modal perturbations that align with observed stall cells.


\subsection{Base configuration and governing equations}

\begin{figure}
    \centering
    \includegraphics[width=0.7\linewidth]{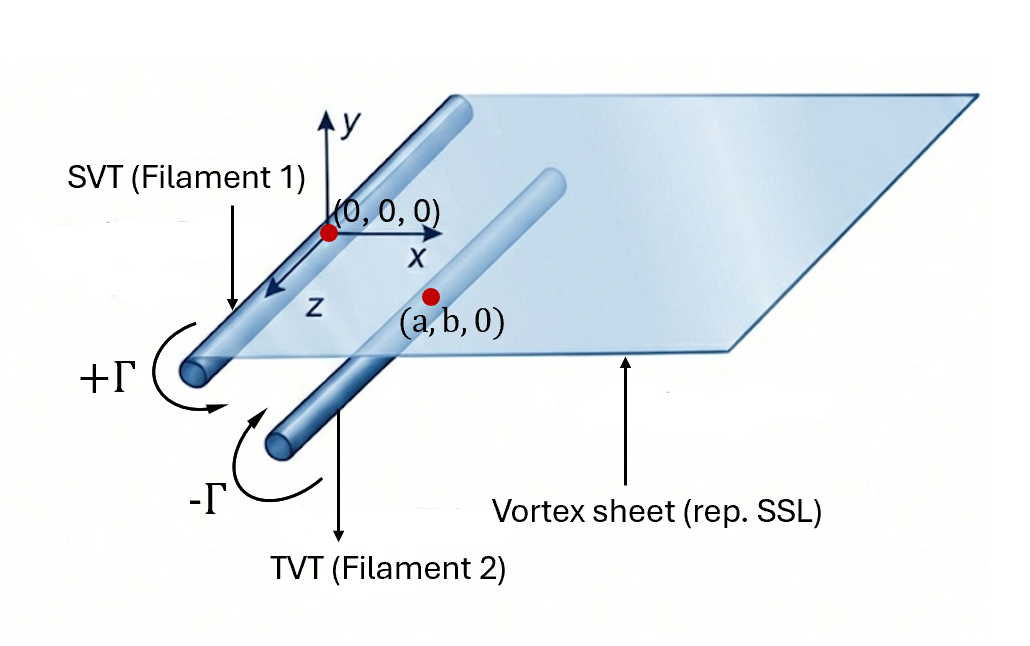}
    \caption{Schematic of the experimental model setup. Key features include the Separation Vortex Tube (SVT), the Trailing edge Vortex Tube (TVT), and the Separated Shear Layer (SSL).}
    \label{fig:Schematic_diagram_vortex}
\end{figure}

To model the essential vortex structures in airfoil stall, we abstract the flow as follows. A schematic diagram of the model setup is given in the Figure \ref{fig:Schematic_diagram_vortex}. The separation vortex tube (denoted filament 1) forms due to boundary layer detachment on the suction side, rolling up into a coherent structure with circulation \(\Gamma > 0\). It is positioned along the spanwise direction from \(z = -s/2\) to \(z = s/2\) at \((x_1, y_1) = (0, 0)\), where \(s\) represents the finite spanwise length, chosen to reflect realistic airfoil spans or computational domain sizes. The trailing edge vortex tube (filament 2) arises from the pressure side flow at the trailing edge, with circulation \(-\Gamma\) to model counter-rotation, a common feature in wake vortex pairs. It is placed at \((x_2, y_2) = (a, b)\), where \(a\) streamwise separation and \(b\) is a small vertical offset, reflecting the geometric asymmetry in airfoil flows; typical values are \(a \sim 1\) (normalised chord) and \(b \ll a\) to ensure close proximity for instability growth.

\noindent The separation shear layer is represented as a vortex sheet attached to filament 1 at \(x = 0\), extending semi-infinitely in \(x > 0\) (downstream) and infinitely in \(z\) for analytical tractability, though finite in practice due to roll-up. The sheet has uniform strength \(\gamma\) (circulation per unit length in \(x\)), selected such that the integrated strength over an effective roll-up length matches \(\Gamma\), i.e., \(\int_0^{l_r} \gamma \, dx = \Gamma\), where \(l_r\) is derived from Kutta-Joukowski conditions or empirical separation lengths. This sheet lies initially in the \(x\)-\(z\) plane at \(y = 0\), with surface vorticity vector \(\boldsymbol{\gamma} = \gamma \mathbf{e}_z\), inducing a tangential velocity jump consistent with separated flows.

\noindent The governing equations are the incompressible Euler equations:
\begin{equation}
\frac{\partial \mathbf{u}}{\partial t} + (\mathbf{u} \cdot \nabla) \mathbf{u} = -\nabla \left( \frac{p}{\rho} \right), \quad \nabla \cdot \mathbf{u} = 0,
\label{eq:euler}
\end{equation}
where \(\mathbf{u}\) is velocity vector, \(p\) is pressure, and \(\rho\) is constant density. Taking the curl of equation \eqref{eq:euler} yields the vorticity transport equation:
\begin{equation}
\frac{D \boldsymbol{\Omega}}{Dt} = (\boldsymbol{\Omega} \cdot \nabla) \mathbf{u},
\label{eq:vorticity_transport}
\end{equation}
with \(\boldsymbol{\Omega} = \nabla \times \mathbf{u}\). In the inviscid limit, vorticity is conserved along fluid particles, justifying concentration on filaments and sheets. The velocity is recovered via the Biot-Savart law:
\begin{equation}
\mathbf{u}(\mathbf{r}) = \frac{1}{4\pi} \int_V \frac{\boldsymbol{\Omega}(\mathbf{r}') \times (\mathbf{r} - \mathbf{r}')}{|\mathbf{r} - \mathbf{r}'|^3} \, dV',
\label{eq:biot_savart}
\end{equation}
chosen over potential formulations for its direct handling of concentrated vorticity. Here $\textbf{r} = (x, y, z)$ denotes the position vector of points in space. This integral decomposes into filament (line) and sheet (surface) contributions, as detailed next.

\subsection{Induced velocities from finite-length vortex filaments and the sheet}

For finite filaments, the Biot-Savart integral \eqref{eq:biot_savart} reduces to a line integral along the path \(C\):
\begin{equation}
\mathbf{u}_\text{fil}(\mathbf{r}) = \frac{\Gamma}{4\pi} \int_C \frac{d\mathbf{l}' \times (\mathbf{r} - \mathbf{r}')}{|\mathbf{r} - \mathbf{r}'|^3}.
\label{eq:biot_savart_filament}
\end{equation}

\noindent For a straight segment from \(\mathbf{A} = (x', y', z_A)\) to \(\mathbf{B} = (x', y', z_B)\) with \(|z_B - z_A| = \Delta z\), we parameterize \(d\mathbf{l}' = dz' \mathbf{e}_z\). Let \(\mathbf{r} = (x, y, z)\), \(\mathbf{d}_\perp = (x - x', y - y', 0)\), \(d = |\mathbf{d}_\perp|\), upper limit \(u = z_B - z\), lower \(l = z_A - z\). 

The key integral in equation \eqref{eq:biot_savart_filament} requires careful evaluation:
\begin{equation}
\int \frac{dz'}{(d^2 + (z - z')^2)^{3/2}} = \left. \frac{z' - z}{d^2 \sqrt{d^2 + (z' - z)^2}} \right|_l^u = I = \frac{u}{d^2 \sqrt{d^2 + u^2}} - \frac{l}{d^2 \sqrt{d^2 + l^2}}.
\label{eq:integral_I}
\end{equation}

\noindent The derivation of \eqref{eq:integral_I} proceeds by substitution: let \(\zeta = z' - z\), then \(d\zeta = dz'\), and we have:
\begin{equation}
\int \frac{d\zeta}{(d^2 + \zeta^2)^{3/2}} = \frac{\zeta}{d^2 \sqrt{d^2 + \zeta^2}} + C.
\end{equation}

\noindent The cross-product \(d\mathbf{l}' \times (\mathbf{r} - \mathbf{r}') = dz' [(y - y') \mathbf{e}_x - (x - x') \mathbf{e}_y]\) yields the velocity components:
\begin{equation}
u^* = \frac{\Gamma}{4\pi} (y - y') I, \quad v^* = -\frac{\Gamma}{4\pi} (x - x') I, \quad w^* = 0.
\label{eq:velocity_components}
\end{equation}

\noindent This form is chosen for its exactness for straight segments, enabling discretization of curved filaments. Finite length introduces end effects (tapering induction at tips), crucial for realistic stall cells with bounded span; infinite approximations overestimate uniform induction.

\noindent For the full filament, we sum over segments. Mutual induction (e.g., on filament 1 from 2) uses \(\Gamma \to -\Gamma\) due to the opposite circulation.

\noindent For the vortex sheet, the surface Biot-Savart integral becomes:
\begin{equation}
\mathbf{u}_\text{sheet}(\mathbf{r}) = \frac{1}{4\pi} \int_S \frac{\boldsymbol{\gamma}(\mathbf{r}') \times (\mathbf{r} - \mathbf{r}')}{|\mathbf{r} - \mathbf{r}'|^3} \, dS'.
\label{eq:biot_savart_sheet}
\end{equation}

\noindent For a flat, infinite uniform sheet with \(\boldsymbol{\gamma} = \gamma \mathbf{e}_z\), the analytical solution gives:
\begin{equation}
\mathbf{u} = \frac{\gamma}{2} \operatorname{sign}(y) \mathbf{e}_x,
\label{eq:sheet_velocity}
\end{equation}
derived via symmetry and Cauchy principal value integration. This represents the velocity jump characteristic of separated flows:
\begin{equation}
[\mathbf{u}]_{y=0^-}^{y=0^+} = \gamma \mathbf{e}_x.
\end{equation}

\noindent The infinite extent in \(z\) simplifies the base state but is perturbed later. The derivation of \eqref{eq:sheet_velocity} uses the fact that for an infinite sheet, contributions from opposite sides of any point cancel in the \(y\) and \(z\) directions, leaving only the \(x\)-component.

\noindent In the base (unperturbed) state, filaments translate due to mutual induction (downward for counter-rotating pairs) plus sheet jump, setting the stage for instability. The base velocity field is:
\begin{equation}
\mathbf{u}_0 = \mathbf{u}_{\text{fil},1} + \mathbf{u}_{\text{fil},2} + \mathbf{u}_{\text{sheet}}.
\label{eq:base_velocity}
\end{equation}

\subsection{Perturbation analysis for Crow-like instability in finite tubes}

To capture bending instabilities, we introduce small perturbations to the filament positions. For finite spanwise length \(s\), we choose a sinusoidal mode fitting pinned boundary conditions (zero displacement at ends, mimicking fixed airfoil tips or domain boundaries):
\begin{align}
\mathbf{r}_1(z, t) &= \left( \xi_1(t) \sin\left(\frac{\pi z}{s}\right), \eta_1(t) \sin\left(\frac{\pi z}{s}\right), z \right), \label{eq:r1_perturb}\\
\mathbf{r}_2(z, t) &= \left( a + \xi_2(t) \sin\left(\frac{\pi z}{s}\right), b + \eta_2(t) \sin\left(\frac{\pi z}{s}\right), z \right). \label{eq:r2_perturb}
\end{align}

\noindent This fundamental mode with wavelength \(\lambda = 2s\) is selected. Higher harmonics could be included in extensions through Fourier series:
\begin{equation}
\mathbf{r}_i(z,t) = \sum_{n=1}^{\infty} \left( \xi_{i,n}(t) \sin\left(\frac{n\pi z}{s}\right), \eta_{i,n}(t) \sin\left(\frac{n\pi z}{s}\right), z \right).
\label{eq:shape of vortex tube}
\end{equation}

\noindent The perturbation amplitudes satisfy \(\xi_{1,2}, \eta_{1,2} \ll b\) to ensure validity of linear analysis.

\noindent Filament motion follows Kelvin's theorem, stating that vortex lines move with the fluid:
\begin{equation}
\frac{\partial \mathbf{r}}{\partial t} = \mathbf{u}(\mathbf{r}),
\label{eq:kelvin_theorem}
\end{equation}
where we neglect self-induction for long waves (\(k b \ll 1\), where \(k = \pi / s\)), as short-wave self-induction is logarithmic in core size and stabilized.

\noindent To analyse stability, we linearize \(\mathbf{v}\) around base positions. Let \(\mathbf{r}_i = \mathbf{r}_{i,0} + \delta \mathbf{r}_i\), where \(\mathbf{r}_{1,0} = (0, 0, z)\) and \(\mathbf{r}_{2,0} = (a, b, z)\). The velocity perturbation is:
\begin{equation}
\delta \mathbf{u} = \frac{\partial \mathbf{u}}{\partial x} \delta x + \frac{\partial \mathbf{u}}{\partial y} \delta y + \mathcal{O}(\delta^2).
\label{eq:velocity_linearization}
\end{equation}

\noindent Computing the velocity derivatives from equations \eqref{eq:velocity_components} and \eqref{eq:integral_I}, we find for the interaction between filaments:
\begin{align}
\frac{\partial u_{1x}}{\partial y_2} &= \frac{\Gamma}{4\pi} \frac{\partial}{\partial y_2} \left[ (y_1 - y_2) I(x_1-x_2, y_1-y_2, z_1-z_2) \right], \\
&= \frac{\Gamma}{4\pi} \left[ -I + (y_1-y_2) \frac{\partial I}{\partial y_2} \right].
\end{align}

\noindent For large \(s/b\) (long tubes relative to separation), averaging over \(z\) and retaining leading-order terms yields the interaction matrix (refer appendix \ref{appex:Detailed derivation of the interaction matrix} for detailed derivation):
\begin{equation}
\mathbf{M} \approx \frac{\Gamma}{2\pi b^2} \begin{pmatrix} 0 & 0 & 0 & 1 \\ 0 & 0 & -1 & 0 \\ 0 & -1 & 0 & 0 \\ 1 & 0 & 0 & 0 \end{pmatrix}.
\label{eq:interaction_matrix}
\end{equation}

\noindent The dynamics of perturbation amplitudes follow:
\begin{equation}
\frac{d}{dt} \begin{pmatrix} \xi_1 \\ \eta_1 \\ \xi_2 \\ \eta_2 \end{pmatrix} = \mathbf{M} \begin{pmatrix} \xi_1 \\ \eta_1 \\ \xi_2 \\ \eta_2 \end{pmatrix}.
\label{eq:perturbation_dynamics}
\end{equation}

\noindent The characteristic equation \(\det(\mathbf{M} - \sigma \mathbf{I}) = 0\) can be expanded as:

\begin{equation}
\left(\sigma^2-\left(\frac{\Gamma}{2\pi b^2}\right)^2\right) = 0
\label{eq:characteristic_eq}
\end{equation}


\noindent giving eigenvalues:
\begin{equation}
\sigma_{1,2} = \frac{\Gamma}{2\pi b^2}, \quad \sigma_{3,4} = - \frac{\Gamma}{2\pi b^2}
\label{eq:eigenvalues}
\end{equation}

\noindent The positive real eigenvalue indicates unstable symmetric bending with growth rate:
\begin{equation}
\sigma_{\text{growth}} = \frac{\Gamma}{2\pi b^2}.
\label{eq:growth_rate}
\end{equation}

\noindent The linear stability analysis reveals the onset mechanism of the Crow-like instability, but predicts unbounded exponential growth that cannot persist in physical systems. To determine the finite-amplitude saturated state that corresponds to observed stall cells, we now develop a weakly nonlinear theory.

\subsection{Weakly nonlinear analysis and amplitude saturation}

The linear stability analysis of previous section predicts exponential growth of perturbations, which cannot persist indefinitely. To determine the finite-amplitude saturated state, we develop a weakly nonlinear theory using the method of multiple scales. This approach systematically accounts for nonlinear interactions while remaining analytically tractable.

\paragraph{Full nonlinear equations of motion.}

The motion of each vortex filament is governed by the Biot-Savart law. For filament $i$ at position $\mathbf{r}_i(z,t)$, the velocity is the sum of mutual induction from filament $j \neq i$ and self-induction:
\begin{equation}
\frac{\partial \mathbf{r}_i}{\partial t} = \mathbf{u}_{j \to i}(\mathbf{r}_i) + \mathbf{u}_{\text{self},i}(\mathbf{r}_i).
\label{eq:full_motion}
\end{equation}

\noindent For two filaments with circulations $\Gamma_1 = \Gamma$ and $\Gamma_2 = -\Gamma$, the mutual induction on filament 1 from filament 2 is:
\begin{equation}
\mathbf{u}_{2 \to 1}(z) = \frac{-\Gamma}{4\pi} \int_{-s/2}^{s/2} \frac{\mathbf{t}_2(z') \times \left[\mathbf{r}_1(z) - \mathbf{r}_2(z')\right]}{\left|\mathbf{r}_1(z) - \mathbf{r}_2(z')\right|^3} \, dz',
\label{eq:mutual_induction_full}
\end{equation}
where $\mathbf{t}_2(z') = \partial \mathbf{r}_2 / \partial z'$ is the unit tangent vector (to leading order) along filament 2.

\noindent The self-induced velocity arises from local curvature and is given by the Local Induction Approximation (LIA):
\begin{equation}
\mathbf{u}_{\text{self},i} = \frac{\Gamma_i}{4\pi} \ln\left(\frac{L}{a_c}\right) \kappa_i \, \hat{\mathbf{b}}_i,
\label{eq:self_induction}
\end{equation}
where $\kappa_i$ is the local curvature, $\hat{\mathbf{b}}_i$ is the binormal vector, $a_c$ is the vortex core radius, and $L$ is a characteristic length scale (typically $\sim 1/k$ for wavenumber $k$). For long-wavelength perturbations where $k a_c \ll 1$, the logarithmic factor $\beta \equiv \ln(L/a_c) \sim \ln(1/ka_c)$ is of order unity to ten.

\paragraph{Perturbation expansion.}

We parameterize the filament positions as:
\begin{align}
\mathbf{r}_1(z,t) &= \left( X_1(z,t), \, Y_1(z,t), \, z \right), \\
\mathbf{r}_2(z,t) &= \left( a + X_2(z,t), \, b + Y_2(z,t), \, z \right),
\end{align}
where $(X_i, Y_i)$ represent displacements from the base configuration. We introduce a small parameter $\epsilon \ll 1$ characterizing the perturbation amplitude relative to the separation:
\begin{equation}
\frac{|X_i|, |Y_i|}{b} \sim \epsilon.
\end{equation}

\noindent The displacement fields are expanded as:
\begin{equation}
\begin{pmatrix} X_i \\ Y_i \end{pmatrix} = \epsilon \begin{pmatrix} X_i^{(1)} \\ Y_i^{(1)} \end{pmatrix} + \epsilon^2 \begin{pmatrix} X_i^{(2)} \\ Y_i^{(2)} \end{pmatrix} + \epsilon^3 \begin{pmatrix} X_i^{(3)} \\ Y_i^{(3)} \end{pmatrix} + \mathcal{O}(\epsilon^4).
\label{eq:amplitude_expansion}
\end{equation}

\noindent We employ the method of multiple scales, introducing:
\begin{equation}
T_0 = t, \quad T_2 = \epsilon^2 t,
\label{eq:time_scales}
\end{equation}
where $T_0$ is the fast time scale of linear oscillations/growth and $T_2$ is the slow time scale on which amplitude modulation occurs. The time derivative becomes:
\begin{equation}
\frac{\partial}{\partial t} = \frac{\partial}{\partial T_0} + \epsilon^2 \frac{\partial}{\partial T_2} + \mathcal{O}(\epsilon^4).
\label{eq:time_derivative}
\end{equation}

\noindent Note, the $\mathcal{O}(\epsilon)$ slow time $T_1 = \epsilon t$ does not appear because the system possesses a reflection symmetry $(X_i, Y_i) \to -(X_i, Y_i)$, which eliminates quadratic nonlinearities in the amplitude equation.

\paragraph{Expansion of mutual induction.}

To obtain the nonlinear terms, we expand the mutual induction \eqref{eq:mutual_induction_full} in powers of the displacements. Define the separation vector:
\begin{equation}
\boldsymbol{\Delta}(z, z') = \mathbf{r}_1(z) - \mathbf{r}_2(z') = \begin{pmatrix} X_1(z) - X_2(z') - a \\ Y_1(z) - Y_2(z') - b \\ z - z' \end{pmatrix}.
\end{equation}

\noindent The base separation (when $X_i = Y_i = 0$) is:
\begin{equation}
\boldsymbol{\Delta}_0(z, z') = \begin{pmatrix} -a \\ -b \\ z - z' \end{pmatrix}, \quad |\boldsymbol{\Delta}_0|^2 = a^2 + b^2 + (z-z')^2 \equiv R_0^2.
\end{equation}

\noindent The perturbation to the separation is:
\begin{equation}
\delta\boldsymbol{\Delta} = \begin{pmatrix} X_1(z) - X_2(z') \\ Y_1(z) - Y_2(z') \\ 0 \end{pmatrix} \equiv \begin{pmatrix} \delta_x \\ \delta_y \\ 0 \end{pmatrix}.
\end{equation}

\noindent We expand $|\boldsymbol{\Delta}|^{-3}$ using:
\begin{equation}
|\boldsymbol{\Delta}|^2 = R_0^2 + 2(\boldsymbol{\Delta}_0 \cdot \delta\boldsymbol{\Delta}) + |\delta\boldsymbol{\Delta}|^2 = R_0^2 \left(1 + 2\chi + \chi_2\right),
\end{equation}
where:
\begin{equation}
\chi = \frac{\boldsymbol{\Delta}_0 \cdot \delta\boldsymbol{\Delta}}{R_0^2} = \frac{-a\delta_x - b\delta_y}{R_0^2}, \quad \chi_2 = \frac{|\delta\boldsymbol{\Delta}|^2}{R_0^2} = \frac{\delta_x^2 + \delta_y^2}{R_0^2}.
\end{equation}

\noindent Expanding to third order:
\begin{align}
|\boldsymbol{\Delta}|^{-3} &= R_0^{-3} \left(1 + 2\chi + \chi_2\right)^{-3/2} \nonumber \\
&= R_0^{-3} \left[ 1 - 3\chi + \frac{15}{2}\chi^2 - \frac{3}{2}\chi_2 - \frac{35}{2}\chi^3 + \frac{15}{2}\chi\chi_2 + \mathcal{O}(\epsilon^4) \right].
\label{eq:kernel_expansion}
\end{align}

\noindent Similarly, the tangent vector on filament 2 expands as:
\begin{equation}
\mathbf{t}_2(z') = \begin{pmatrix} \partial_{z'} X_2 \\ \partial_{z'} Y_2 \\ 1 \end{pmatrix} = \begin{pmatrix} 0 \\ 0 \\ 1 \end{pmatrix} + \epsilon \begin{pmatrix} \partial_{z'} X_2^{(1)} \\ \partial_{z'} Y_2^{(1)} \\ 0 \end{pmatrix} + \mathcal{O}(\epsilon^2).
\end{equation}

\noindent Substituting these expansions into \eqref{eq:mutual_induction_full} and collecting terms by order in $\epsilon$ yields the systematic expansion of the equations of motion.

\paragraph{Order $\mathcal{O}(\epsilon)$: Linear problem.}

At leading order, we recover the linear system. For sinusoidal perturbations with wavenumber $k = \pi/s$:
\begin{equation}
\begin{pmatrix} X_i^{(1)} \\ Y_i^{(1)} \end{pmatrix} = \begin{pmatrix} \xi_i(T_0, T_2) \\ \eta_i(T_0, T_2) \end{pmatrix} \sin(kz),
\end{equation}
the equations become:
\begin{equation}
\frac{\partial}{\partial T_0} \begin{pmatrix} \xi_1 \\ \eta_1 \\ \xi_2 \\ \eta_2 \end{pmatrix} = \mathbf{M} \begin{pmatrix} \xi_1 \\ \eta_1 \\ \xi_2 \\ \eta_2 \end{pmatrix},
\label{eq:linear_system}
\end{equation}
where $\mathbf{M}$ is the interaction matrix derived previously:
\begin{equation}
\mathbf{M} = \frac{\Gamma}{2\pi b^2} \begin{pmatrix} 0 & 0 & 0 & 1 \\ 0 & 0 & -1 & 0 \\ 0 & -1 & 0 & 0 \\ 1 & 0 & 0 & 0 \end{pmatrix} \equiv \sigma_0 \begin{pmatrix} 0 & 0 & 0 & 1 \\ 0 & 0 & -1 & 0 \\ 0 & -1 & 0 & 0 \\ 1 & 0 & 0 & 0 \end{pmatrix},
\label{eq:M_matrix}
\end{equation}
with $\sigma_0 = \Gamma/(2\pi b^2)$ being the characteristic growth rate.

\noindent The eigenvalues are $\sigma = \pm \sigma_0$, each with multiplicity 2. The eigenvectors for the unstable eigenvalue $\sigma = +\sigma_0$ are:
\begin{equation}
\mathbf{v}_1 = \begin{pmatrix} 1 \\ 0 \\ 0 \\ 1 \end{pmatrix}, \quad \mathbf{v}_2 = \begin{pmatrix} 0 \\ 1 \\ -1 \\ 0 \end{pmatrix}.
\label{eq:unstable_eigenvectors}
\end{equation}

\noindent For the stable eigenvalue $\sigma = -\sigma_0$:
\begin{equation}
\mathbf{v}_3 = \begin{pmatrix} 1 \\ 0 \\ 0 \\ -1 \end{pmatrix}, \quad \mathbf{v}_4 = \begin{pmatrix} 0 \\ 1 \\ 1 \\ 0 \end{pmatrix}.
\label{eq:stable_eigenvectors}
\end{equation}

\noindent The general solution at $\mathcal{O}(\epsilon)$ is:
\begin{equation}
\mathbf{q}^{(1)} = A_1(T_2) e^{\sigma_0 T_0} \mathbf{v}_1 + A_2(T_2) e^{\sigma_0 T_0} \mathbf{v}_2 + \text{(decaying modes)},
\label{eq:linear_solution}
\end{equation}
where $A_1(T_2)$ and $A_2(T_2)$ are slowly-varying complex amplitudes to be determined by solvability conditions at higher order.

\paragraph{Order $\mathcal{O}(\epsilon^2)$: Quadratic corrections.}

At second order, the equations take the form:
\begin{equation}
\frac{\partial \mathbf{q}^{(2)}}{\partial T_0} - \mathbf{M} \mathbf{q}^{(2)} = \mathbf{N}_2(\mathbf{q}^{(1)}, \mathbf{q}^{(1)}),
\label{eq:second_order}
\end{equation}
where $\mathbf{N}_2$ represents quadratic nonlinearities arising from the expansion \eqref{eq:kernel_expansion}.

\noindent Explicit computation shows that $\mathbf{N}_2$ involves terms proportional to $\sin^2(kz) = \frac{1}{2}(1 - \cos(2kz))$. The $\cos(2kz)$ component generates a second harmonic, while the constant term produces a mean-flow correction. Crucially, neither of these resonates with the fundamental mode $\sin(kz)$, so no solvability condition arises at this order.

\noindent The second-order solution is:
\begin{equation}
\mathbf{q}^{(2)} = \mathbf{q}^{(2)}_0 + \mathbf{q}^{(2)}_{2k} \cos(2kz),
\end{equation}
representing mean-flow distortion and second-harmonic generation.

\paragraph{Order $\mathcal{O}(\epsilon^3)$: Cubic nonlinearity and amplitude equation.}

At third order:
\begin{equation}
\frac{\partial \mathbf{q}^{(3)}}{\partial T_0} - \mathbf{M} \mathbf{q}^{(3)} = -\frac{\partial \mathbf{q}^{(1)}}{\partial T_2} + \mathbf{N}_3(\mathbf{q}^{(1)}, \mathbf{q}^{(1)}, \mathbf{q}^{(1)}) + \mathbf{N}_2(\mathbf{q}^{(1)}, \mathbf{q}^{(2)}) + \mathbf{S}(\mathbf{q}^{(1)}),
\label{eq:third_order}
\end{equation}
where $\mathbf{N}_3$ represents cubic nonlinearities from \eqref{eq:kernel_expansion}, the $\mathbf{N}_2(\mathbf{q}^{(1)}, \mathbf{q}^{(2)})$ term captures the interaction between first and second-order solutions, and $\mathbf{S}$ represents self-induction contributions.

\noindent The cubic terms produce components proportional to $\sin^3(kz) = \frac{3}{4}\sin(kz) - \frac{1}{4}\sin(3kz)$. The $\sin(kz)$ part resonates with the linear eigenmode, generating secular growth that must be eliminated for a uniformly valid expansion.

\paragraph{Solvability condition and Stuart-Landau equation.}

The solvability condition requires that the right-hand side of \eqref{eq:third_order} be orthogonal to the null space of the adjoint operator $(\partial_{T_0} - \mathbf{M}^T)$. The adjoint eigenvectors for eigenvalue $+\sigma_0$ are:
\begin{equation}
\mathbf{w}_1 = \begin{pmatrix} 0 \\ 0 \\ 0 \\ 1 \end{pmatrix}, \quad \mathbf{w}_2 = \begin{pmatrix} 0 \\ -1 \\ 0 \\ 0 \end{pmatrix}.
\end{equation}

\noindent Taking the inner product with $\mathbf{w}_1$ and $\mathbf{w}_2$, we obtain the amplitude equations:
\begin{align}
\frac{dA_1}{dT_2} &= -\ell_{11} |A_1|^2 A_1 - \ell_{12} |A_2|^2 A_1 - \mu_{12} A_1^* A_2^2, \label{eq:A1_evolution} \\
\frac{dA_2}{dT_2} &= -\ell_{22} |A_2|^2 A_2 - \ell_{21} |A_1|^2 A_2 - \mu_{21} A_2^* A_1^2, \label{eq:A2_evolution}
\end{align}
where the Landau coefficients $\ell_{ij}$ and coupling coefficients $\mu_{ij}$ are real and determined by the cubic nonlinearities.

\paragraph{Explicit calculation of Landau coefficients.}

The dominant contribution to the Landau coefficients comes from the geometric nonlinearity in the Biot-Savart kernel. From the expansion \eqref{eq:kernel_expansion}, the cubic terms proportional to $\chi^3$ and $\chi \chi_2$ contribute.

\noindent For the symmetric mode $\mathbf{v}_1$ (where $\xi_1 = \xi_2 \equiv \xi$ and $\eta_1 = \eta_2 = 0$), the perturbation to the separation in the $x$-direction is:
\begin{equation}
\delta_x = X_1 - X_2 = \xi \sin(kz) - \xi \sin(kz') = \xi \left[\sin(kz) - \sin(kz')\right].
\end{equation}

\noindent For nearby points $z' \approx z$, this is small, but the cumulative effect integrated over the filament produces finite corrections.

\noindent After extensive algebra (detailed in Appendix \ref{appex:Landau_coefficient}), the self-interaction Landau coefficient is:
\begin{equation}
\ell_{11} = \frac{3\Gamma}{4\pi b^4} \left( \frac{a^2 + b^2}{b^2} \right) \mathcal{I}(kb),
\label{eq:Landau_coefficient}
\end{equation}
where the dimensionless integral $\mathcal{I}(kb)$ is:
\begin{equation}
\mathcal{I}(kb) = \int_0^\infty \frac{u^2 (1 + u^2)^{-5/2}}{\left[1 + (kb)^2(1+u^2)\right]} \, du.
\label{eq:I_integral}
\end{equation}

\noindent For long waves ($kb \ll 1$), this integral evaluates to:
\begin{equation}
\mathcal{I}(kb) \approx \frac{2}{3} - \frac{4}{15}(kb)^2 + \mathcal{O}(kb)^4,
\end{equation}
giving:
\begin{equation}
\ell_{11} \approx \frac{\Gamma}{2\pi b^4} \left( \frac{a^2 + b^2}{b^2} \right) \quad \text{for } kb \ll 1.
\label{eq:Landau_long_wave}
\end{equation}

\noindent The self-induction contribution from \eqref{eq:self_induction} adds a stabilizing term:
\begin{equation}
\ell_{\text{self}} = \frac{\Gamma \beta k^2}{4\pi},
\label{eq:Landau_self}
\end{equation}
where $\beta = \ln(1/ka_c)$.

\noindent The total Landau coefficient is:
\begin{equation}
\ell = \ell_{11} + \ell_{\text{self}} = \frac{\Gamma}{2\pi b^4} \left( \frac{a^2 + b^2}{b^2} \right) + \frac{\Gamma \beta k^2}{4\pi}.
\label{eq:Landau_total}
\end{equation}

\paragraph{Single-mode Stuart-Landau equation.}

For a single dominant mode (say $A_1$ with $A_2 = 0$), the amplitude equation \eqref{eq:A1_evolution} simplifies to the Stuart-Landau form. Reverting to the original time variable $t$ and defining the physical amplitude $A = \epsilon A_1$:
\begin{equation}
\frac{dA}{dt} = \sigma_0 A - \ell |A|^2 A,
\label{eq:Stuart_Landau}
\end{equation}
where $\sigma_0 = \Gamma/(2\pi b^2)$ is the linear growth rate.

\paragraph{Saturation amplitude and time scale.}

The Stuart-Landau equation \eqref{eq:Stuart_Landau} admits the steady-state solution:
\begin{equation}
|A|_{\text{sat}} = \sqrt{\frac{\sigma_0}{\ell}}.
\label{eq:saturation_amplitude}
\end{equation}

\noindent Substituting the expressions for $\sigma_0$ and $\ell$:
\begin{equation}
|A|_{\text{sat}} = b \left[ \frac{b^2}{a^2 + b^2 + \frac{\beta (kb)^2 b^4}{2}} \right]^{1/2}.
\label{eq:saturation_explicit}
\end{equation}

\noindent For the typical case where  $kb \ll 1$ (long wavelength compared to separation):
\begin{equation}
|A|_{\text{sat}} \approx \frac{b^2}{\sqrt{a^2 + b^2}} \sim \mathcal{O}(b).
\label{eq:saturation_scaling}
\end{equation}

\noindent This confirms that the saturated bending amplitude scales with the filament separation $b$, validating the weakly nonlinear assumption that saturation occurs when perturbations become comparable to the base-state geometry.

\noindent The approach to saturation follows:
\begin{equation}
|A(t)|^2 = \frac{|A|_{\text{sat}}^2}{1 + \left(\frac{|A|_{\text{sat}}^2}{|A_0|^2} - 1\right) e^{-2\sigma_0 t}},
\label{eq:amplitude_evolution}
\end{equation}
where $A_0$ is the initial amplitude. The saturation time scale is:
\begin{equation}
\tau_{\text{sat}} \sim \frac{1}{2\sigma_0} \ln\left(\frac{|A|_{\text{sat}}}{|A_0|}\right) = \frac{\pi b^2}{\Gamma} \ln\left(\frac{|A|_{\text{sat}}}{|A_0|}\right).
\label{eq:saturation_time}
\end{equation}

\paragraph{Physical interpretation and regime of validity.}

The saturation mechanism can be understood physically as follows:

\begin{enumerate}
    \item \textbf{Geometric feedback}: As the filaments bend with amplitude $|A|$, the local separation varies along the span. At locations where filaments approach (separation $< b$), mutual induction intensifies; where they recede (separation $> b$), it weakens. This modulation of the interaction strength provides a self-limiting feedback.
    
    \item \textbf{Self-induction stabilization}: The bending increases local curvature $\kappa \sim |A| k^2$, generating self-induced velocities that oppose further bending. This effect, captured by $\ell_{\text{self}}$ in \eqref{eq:Landau_self}, becomes increasingly important for shorter wavelengths.
    
    \item \textbf{Harmonic generation}: Nonlinear interactions transfer energy from the fundamental mode to higher harmonics ($2k$, $3k$, ...), effectively draining energy from the unstable mode.
\end{enumerate}

\noindent The weakly nonlinear theory is valid when:
\begin{equation}
\epsilon = \frac{|A|}{b} \lesssim 0.3,
\label{eq:validity}
\end{equation}
beyond which higher-order terms become significant. For stronger nonlinearity, numerical simulation or asymptotic matching to a fully nonlinear regime is required.

\paragraph{Implications for stall cell wavelength selection.}

The saturated amplitude \eqref{eq:saturation_explicit} depends on the wavenumber $k$ through both the linear growth rate and the Landau coefficient. The most prominent stall cell pattern corresponds to the mode that achieves the largest saturated amplitude.

\noindent Differentiating $|A|_{\text{sat}}^2$ with respect to $k$ and setting to zero:
\begin{equation}
\frac{d|A|_{\text{sat}}^2}{dk} = 0 \implies k_{\text{opt}} \sim \frac{1}{b} \left( \frac{a^2 + b^2}{\beta b^2} \right)^{1/2}.
\label{eq:optimal_k}
\end{equation}

\noindent For $a \sim 0$:
\begin{equation}
k_{\text{opt}} \sim \frac{1}{b}, \quad \lambda_{\text{opt}} = \frac{2\pi}{k_{\text{opt}}} \sim 2\pi b.
\label{eq:optimal_wavelength}
\end{equation}

This predicts that the dominant stall cell wavelength scales with the separation between the vortex tubes, providing a testable relationship between the flow geometry and observed stall cell spacing.

\subsection{Summary of nonlinear theory}

The weakly nonlinear analysis yields the following key results:

\begin{enumerate}
    \item The Crow-like instability saturates at finite amplitude due to geometric nonlinearity in the Biot-Savart interaction and self-induction effects.
    
    \item The saturated amplitude scales as:
    \begin{equation}
    |A|_{\text{sat}} \sim b \cdot f\left(\frac{a}{b}, kb, \beta\right),
    \end{equation}
    where $f$ is an $\mathcal{O}(1)$ function given by \eqref{eq:saturation_explicit}.
    
    \item The saturation time scale is:
    \begin{equation}
    \tau_{\text{sat}} \sim \frac{\pi b^2}{\Gamma} \ln\left(\frac{b}{|A_0|}\right).
    \end{equation}
    
    \item The optimal wavelength for maximum saturated amplitude scales as $\lambda_{\text{opt}} \sim 2\pi b$.
    
    \item The theory predicts that stronger circulation $\Gamma$ leads to faster saturation but the same saturated amplitude, while larger separation $b$ increases both the saturation time and the saturated amplitude.
\end{enumerate}

These predictions provide quantitative benchmarks for comparison with numerical simulations and experimental observations of stall cells.



\subsection{Coupling the vortex sheet deformation to the filament bend}

The sheet dynamics are governed by the 3D Birkhoff-Rott equation, the surface analog of filament motion:
\begin{equation}
\frac{\partial \mathbf{Z}}{\partial t} = \frac{1}{4\pi} \text{P.V.} \int_S \frac{\boldsymbol{\gamma}(\mathbf{p}') \times (\mathbf{Z}(\mathbf{p}, t) - \mathbf{Z}(\mathbf{p}', t))}{|\mathbf{Z}(\mathbf{p}, t) - \mathbf{Z}(\mathbf{p}', t)|^3} \, dP' + \mathbf{u}_\text{ext}(\mathbf{Z}),
\label{eq:birkhoff_rott}
\end{equation}
where \(\mathbf{Z}(p, l, t)\) parametrizes the sheet (\(p\) along \(x\), \(l\) along \(z\)), P.V. denotes the Cauchy principal value (to handle the singularity at \(\mathbf{p}' = \mathbf{p}\)), and \(\mathbf{u}_\text{ext}\) includes filament-induced velocities from equation \eqref{eq:biot_savart_filament}.

\noindent The derivation of equation \eqref{eq:birkhoff_rott} follows from the requirement that the sheet, as a material surface of discontinuity, moves with the average velocity across it:
\begin{equation}
\frac{\partial \mathbf{Z}}{\partial t} = \frac{1}{2}\left[\mathbf{u}^+ + \mathbf{u}^-\right]_{\text{at } \mathbf{Z}},
\end{equation}
ensuring no flow through the sheet.

\noindent The attachment condition enforces continuity between sheet edge and filament:
\begin{equation}
\mathbf{Z}(p=0, l, t) = \mathbf{r}_1(l, t) \quad \text{for } l \in [-s/2, s/2],
\label{eq:attachment}
\end{equation}
coupling sheet edge deformation to filament bend through equations \eqref{eq:r1_perturb}.

\noindent For small deformations, we parameterize the sheet position as:
\begin{equation}
\mathbf{Z}(p, l, t) = (p, h(p, l, t), l),
\label{eq:sheet_param}
\end{equation}
with \(h \ll 1\) representing vertical displacement from the base plane.

\noindent Linearizing the Birkhoff-Rott equation \eqref{eq:birkhoff_rott} about the flat state and neglecting higher-order self-induction for weak curvature, we obtain:
\begin{equation}
\frac{\partial h}{\partial t} = \frac{1}{4\pi} \text{P.V.} \int \frac{\gamma (l' - l)}{[(p-p')^2 + (l-l')^2]^{3/2}} h(p', l', t) \, dp' dl' + u_{y,\text{ext}}.
\label{eq:linearized_br}
\end{equation}

\noindent We assume an exponential decay form consistent with the attachment condition \eqref{eq:attachment}:
\begin{equation}
h(p, l, t) = \eta_1(t) e^{-\alpha p} \cos\left(\frac{\pi l}{s}\right),
\label{eq:h_form}
\end{equation}
where \(\alpha\) is the spatial decay rate. Here we have chosen \(\alpha \sim k = \pi/s\), ensuring evanescent decay over one wavelength. Though, more work needs to be done to find out way to rigorously find a value for \(\alpha\), and it is planned to be done in the future.

\noindent The physical reasoning for equation \eqref{eq:h_form}: 
- At \(p=0\): \(h(0, l, t) = \eta_1(t) \sin(\pi l / s)\), matching filament 1 displacement
- As \(p \to \infty\): \(h \to 0\), recovering flat sheet far downstream
- The sinusoidal \(l\)-dependence propagates the filament's spanwise mode

\noindent Substituting \eqref{eq:h_form} into \eqref{eq:linearized_br} and using the attachment condition time derivative:
\begin{equation}
\frac{\partial h}{\partial t}\bigg|_{p=0} = \frac{d\eta_1}{dt} \sin\left(\frac{\pi l}{s}\right) = \sigma \eta_1 \sin\left(\frac{\pi l}{s}\right),
\end{equation}
provides a consistency check on the decay rate \(\alpha\).

\subsection{Induction of alternating y-vorticity from sheet deformation}

The deformed sheet alters the local geometry fundamentally. The surface position is given by equation \eqref{eq:sheet_param}, and we compute the geometric quantities:

\noindent Surface tangent vectors:
\begin{align}
\mathbf{t}_p &= \frac{\partial \mathbf{Z}}{\partial p} = \left(1, \frac{\partial h}{\partial p}, 0\right), \\
\mathbf{t}_l &= \frac{\partial \mathbf{Z}}{\partial l} = \left(0, \frac{\partial h}{\partial l}, 1\right).
\end{align}

\noindent Surface normal (unnormalised):
\begin{equation}
\mathbf{n} = \mathbf{t}_p \times \mathbf{t}_l = \left(\frac{\partial h}{\partial p}, -1, \frac{\partial h}{\partial l}\right).
\label{eq:surface_normal}
\end{equation}

\noindent To first order in \(h\), the normalised versions are:
\begin{align}
\hat{\mathbf{t}}_p &\approx \left(1, \frac{\partial h}{\partial p}, 0\right), \\
\hat{\mathbf{t}}_l &\approx \left(0, \frac{\partial h}{\partial l}, 1\right), \\
\hat{\mathbf{n}} &\approx \left(\frac{\partial h}{\partial p}, -1, \frac{\partial h}{\partial l}\right).
\end{align}

\noindent The surface vorticity, originally \(\boldsymbol{\gamma} = \gamma \mathbf{e}_z\) in the flat reference frame, must be projected onto the deformed tangent basis. Since vorticity is a material vector in inviscid flow, its magnitude is conserved but its direction follows the surface deformation:
\begin{equation}
\boldsymbol{\gamma}_{\text{loc}} = \gamma \hat{\mathbf{t}}_l \approx \gamma \left(0, \frac{\partial h}{\partial l}, 1\right).
\label{eq:local_vorticity}
\end{equation}

\noindent Thus, the \(y\)-component of surface vorticity is:
\begin{equation}
\gamma_y = \gamma \frac{\partial h}{\partial l}.
\label{eq:gamma_y}
\end{equation}

\noindent The volume-equivalent vorticity field is expressed using the surface delta function:
\begin{equation}
\boldsymbol{\Omega}(\mathbf{r}) = \boldsymbol{\gamma}(\mathbf{Z}) \delta(P),
\end{equation}
where \(\delta(P)\) concentrates vorticity on the sheet surface.

\noindent For small deformations, we expand:
\begin{equation}
\delta(y - h(x,z)) \approx \delta(y) - h(x,z) \delta'(y) + \mathcal{O}(h^2).
\end{equation}

\noindent Retaining the leading term for small \(h\):
\begin{equation}
\Omega_y(\mathbf{r}) \approx \gamma \frac{\partial h}{\partial z} \delta(y).
\label{eq:omega_y}
\end{equation}

\noindent Substituting the form \eqref{eq:h_form} for \(h\):
\begin{align}
\frac{\partial h}{\partial z} &= \eta_1(t) e^{-\alpha x} \frac{\pi}{s} \cos\left(\frac{\pi z}{s}\right), \\
\Omega_y &\approx \gamma \eta_1(t) \frac{\pi}{s} e^{-\alpha x} \cos\left(\frac{\pi z}{s}\right) \delta(y).
\label{eq:omega_y_explicit}
\end{align}

\noindent The cosine function creates alternating sign in \(\Omega_y\) along the span, with period \(2s\). During the instability growth (positive \(\sigma\) from equation \eqref{eq:growth_rate}), we have:
\begin{equation}
\Omega_y \approx \gamma \epsilon \frac{\pi}{s} e^{-\alpha x + \sigma t} \cos\left(\frac{\pi z}{s}\right) \delta(y),
\end{equation}
where \(\epsilon\) is the initial perturbation amplitude.

\noindent This alternating \(y\)-vorticity represents the key physical mechanism: the sheet deformation induced by filament bending creates spanwise-periodic vertical vorticity, absent in the base state.

\subsection{Alternating spanwise velocity induced by y-vorticity}

The \(y\)-component vorticity induces velocity throughout the flow field via the Biot-Savart law. From equation \eqref{eq:biot_savart}, the \(z\)-component of velocity is:
\begin{equation}
w^*(\mathbf{r}) = \frac{1}{4\pi} \int \frac{\Omega_y(\mathbf{r}') (x - x') - \Omega_x(\mathbf{r}') (z - z')}{|\mathbf{r} - \mathbf{r}'|^3} \, dV',
\label{eq:uz_from_omega}
\end{equation}
where the dominant contribution comes from \(\Omega_y\) (as \(\Omega_x\) is negligible in our configuration).

\noindent Given the form \eqref{eq:omega_y_explicit}:
\begin{equation}
\Omega_y = \zeta(x) \cos\left(\frac{\pi z}{s}\right) \delta(y),
\end{equation}
with amplitude function:
\begin{equation}
\zeta(x) = \gamma \epsilon \frac{\pi}{s} e^{-\alpha x + \sigma t}.
\end{equation}

\noindent To solve for the induced velocity, we use the stream function approach. For 2D incompressible flow in the \(x\)-\(z\) plane at fixed \(y\), the stream function \(\psi(x,y,z)\) satisfies:
\begin{align}
u^* &= \frac{\partial \psi}{\partial z}, \\
w^* &= -\frac{\partial \psi}{\partial x}, \\
\nabla^2 \psi &= -\Omega_y.
\label{eq:stream_function}
\end{align}

Fourier decomposing in \(z\):
\begin{equation}
\psi(x,y,z) = \sum_k \hat{\psi}_k(x,y) e^{ikz}, \quad \Omega_y = \sum_k \hat{\Omega}_{y,k}(x,y) e^{ikz}.
\end{equation}

For our cosine distribution, the dominant mode has \(k = \pi/s\):
\begin{equation}
\hat{\Omega}_{y,k} = \frac{\zeta(x)}{2} \delta(y) \quad \text{for } k = \pm\pi/s.
\end{equation}

The Fourier-transformed equation becomes:
\begin{equation}
\left(\frac{\partial^2}{\partial x^2} + \frac{\partial^2}{\partial y^2} - k^2\right) \hat{\psi}_k = -\hat{\Omega}_{y,k}.
\label{eq:helmholtz}
\end{equation}

\noindent For the line source at \(y = 0\), the Green's function solution involves modified Bessel functions. In the far field (\(|y| \gg 1/k\)), the solution decays exponentially:
\begin{equation}
\hat{\psi}_k \sim \frac{1}{2k} \int_{-\infty}^{\infty} \zeta(x') e^{-k|y-y'|} e^{-ik(x-x')} dx' dy'.
\end{equation}

\noindent For our exponentially decaying \(\zeta(x) = \zeta_0 e^{-\alpha x}\) (with \(x > 0\)):
\begin{equation}
\int_0^{\infty} \zeta_0 e^{-\alpha x'} e^{-ik(x-x')} dx' = \frac{\zeta_0 e^{-ikx}}{\alpha - ik} \quad \text{for } \text{Re}(\alpha) > 0.
\end{equation}

\noindent After integrating over the delta function at \(y' = 0\) and taking the real part, the spanwise velocity becomes:
\begin{equation}
w^* \approx \frac{\zeta_0 s}{\pi(\alpha^2 + k^2)^{1/2}} e^{-k|y|} \sin\left(\frac{\pi z}{s} + \phi\right) e^{\sigma t},
\label{eq:uz_final}
\end{equation}
where \(\phi = \arctan(k/\alpha)\) is a phase shift, and \(\zeta_0 = \gamma \epsilon \pi/s\).

\noindent For the typical case \(\alpha \approx k = \pi/s\):
\begin{equation}
w^* \approx \frac{\gamma \epsilon s}{2\pi} e^{-\pi|y|/s} \sin\left(\frac{\pi z}{s} + \frac{\pi}{4}\right) e^{\sigma t}.
\end{equation}

\noindent Key features of this spanwise velocity field:
\begin{itemize}
    \item \textbf{Alternating sign}: The $\sin(\pi z/s)$ factor creates regions of positive and negative $w^*$.
    \item \textbf{Exponential growth}: The $e^{\sigma t}$ factor shows instability-driven amplification.
    \item \textbf{Vertical decay}: The $e^{-\pi |y|/s}$ factor confines the effect near the sheet.
    \item \textbf{Spanwise wavelength}: The pattern repeats with wavelength $2s$, matching the initial perturbation.
\end{itemize}

This oscillatory \(w^*\) field represents the hallmark of stall cells: alternating regions of spanwise flow that create the characteristic cellular pattern. In regions where \(w^* > 0\), fluid moves in the \(+z\) direction, while adjacent regions have \(w^* < 0\). This spanwise redistribution of momentum leads to alternating patterns of flow separation and reattachment along the airfoil span.

\noindent The complete velocity field induced by the deformed vortex structures is:
\begin{equation}
\mathbf{u}_{\text{induced}} = \mathbf{u}_{\text{filaments}} + \mathbf{u}_{\text{sheet}} + \mathbf{u}_{\omega_y},
\end{equation}
where each component contributes to the complex three-dimensional flow structure characteristic of stall cells.

\subsection{Physical mechanism summary and validation}

The complete mechanism for stall cell formation can be summarised as follows:

\begin{itemize}
    \item \textbf{Initial configuration}: Counter-rotating vortex tubes (separation and trailing edge) in proximity, with attached shear layer
    \item \textbf{Crow instability}: Mutual induction drives symmetric bending with growth rate \(\sigma\) (equation \eqref{eq:growth_rate})
    \item \textbf{Sheet deformation}: Attachment condition couples filament bend to sheet shape (equations \eqref{eq:attachment} and \eqref{eq:h_form})
    \item \textbf{Vorticity generation}: Sheet tilting induces alternating \(y\)-vorticity (equation \eqref{eq:omega_y_explicit})
    \item \textbf{Spanwise flow}: The \(y\)-vorticity drives alternating spanwise velocity (equation \eqref{eq:uz_final})
    \item \textbf{Stall cells}: Spanwise flow redistribution creates periodic separation patterns
\end{itemize}


\subsubsection{Saturated flow field quantities}

Having established the saturation amplitude $A_{\text{sat}} = b^2/\sqrt{a^2+b^2}$ from the Stuart-Landau analysis, we now derive the saturated expressions for the observable flow quantities.

\paragraph{Saturated vortex sheet deformation.}

From equation \eqref{eq:h_form}, the vortex sheet deformation is:
\begin{equation}
h(x, z, t) = \eta_1(t) \, e^{-\alpha x} \cos\left(\frac{\pi z}{s}\right),
\end{equation}
where $\alpha = \pi/s$ is the streamwise decay rate. The cosine dependence ensures that the sheet deformation is maximum at the cell boundaries ($z = 0, \pm s$) and zero at the cell centers ($z = \pm s/2$), consistent with the physical requirement that the maximum spanwise velocity occurs on the sheet surface. The bending amplitude $\eta_1(t)$ evolves according to the Stuart-Landau equation and saturates at $\eta_1 \to A_{\text{sat}}$.

Substituting $\eta_1 = A_{\text{sat}} = b^2/\sqrt{a^2 + b^2}$:
\begin{align}
h_{\text{sat}}(x, z) &= A_{\text{sat}} \, e^{-\alpha x} \cos\left(\frac{\pi z}{s}\right) \nonumber \\[1ex]
&= \frac{b^2}{\sqrt{a^2 + b^2}} \, e^{-\pi x/s} \cos\left(\frac{\pi z}{s}\right).
\end{align}

Thus:
\begin{equation}
\boxed{h_{\text{sat}}(x, z) = \frac{b^2}{\sqrt{a^2 + b^2}} \, e^{-\pi x/s} \cos\left(\frac{\pi z}{s}\right).}
\label{eq:h_sat_final}
\end{equation}

The maximum occurs at $x = 0$ and $z = 0$ where $\cos(0) = 1$:
\begin{equation}
h_{\max} = \frac{b^2}{\sqrt{a^2 + b^2}}.
\end{equation}

For $a = 0$:
\begin{equation}
h_{\max} = \frac{b^2}{\sqrt{b^2}} = \frac{b^2}{b} = b 
\end{equation}

\paragraph{Saturated vertical vorticity.}

From equation \eqref{eq:gamma_y}, the $y$-component of vorticity induced by sheet deformation is:
\begin{equation}
\Omega_y = \gamma \frac{\partial h}{\partial z} \, \delta(y),
\end{equation}
where $\gamma$ is the vortex sheet strength.

Computing the partial derivative of $h_{\text{sat}}$ with respect to $z$:
\begin{align}
\frac{\partial h_{\text{sat}}}{\partial z} &= \frac{\partial}{\partial z} \left[ \frac{b^2}{\sqrt{a^2 + b^2}} \, e^{-\pi x/s} \cos\left(\frac{\pi z}{s}\right) \right] \nonumber \\[1ex]
&= \frac{b^2}{\sqrt{a^2 + b^2}} \, e^{-\pi x/s} \cdot \frac{\pi}{s} \sin\left(\frac{\pi z}{s}\right) \nonumber \\[1ex]
&= \frac{\pi b^2}{s\sqrt{a^2 + b^2}} \, e^{-\pi x/s} \sin\left(\frac{\pi z}{s}\right).
\end{align}

Multiplying by $\gamma$ and including the delta function:
\begin{align}
\Omega_{y,\text{sat}} &= \gamma \cdot \frac{\pi b^2}{s\sqrt{a^2 + b^2}} \, e^{-\pi x/s} \sin\left(\frac{\pi z}{s}\right) \cdot \delta(y).
\end{align}

Thus:
\begin{equation}
\boxed{\Omega_{y,\text{sat}}(x, z) = \frac{\gamma \pi b^2}{s\sqrt{a^2 + b^2}} \, e^{-\pi x/s} \sin\left(\frac{\pi z}{s}\right) \, \delta(y).}
\label{eq:Omega_y_sat_final}
\end{equation}

The maximum magnitude occurs at $x = 0$ and $z = \pm s/2$ where $|\sin(\pm\pi/2)| = 1$:
\begin{equation}
|\Omega_{y,\max}| = \frac{\gamma \pi b^2}{s\sqrt{a^2 + b^2}}.
\end{equation}

For $a = 0$:
\begin{equation}
|\Omega_{y,\max}| = \frac{\gamma \pi b^2}{sb} = \frac{\gamma \pi b}{ s}.
\end{equation}

The sinusoidal dependence ensures that the vorticity is zero at the cell boundaries ($z = 0, \pm s$) and reaches extrema at the cell centers ($z = \pm s/2$), with $\Omega_y > 0$ for $z > 0$ and $\Omega_y < 0$ for $z < 0$.

\paragraph{Saturated spanwise velocity.}

From equation \eqref{eq:uz_final}, the spanwise velocity induced by the $y$-vorticity is:
\begin{equation}
w^*(x, y, z) = \frac{\gamma \eta_1 s}{2\pi} \, e^{-\alpha x} \, e^{-\pi|y|/s} \sin\left(\frac{\pi z}{s}\right),
\end{equation}
where the sine dependence arises from the Biot-Savart induction, maintaining the $\pi/2$ phase relationship with the sheet deformation.

Substituting $\eta_1 = A_{\text{sat}} = b^2/\sqrt{a^2 + b^2}$ and $\alpha = \pi/s$:
\begin{align}
w^*_{\text{sat}} &= \frac{\gamma s}{2\pi} \cdot \frac{b^2}{\sqrt{a^2 + b^2}} \cdot e^{-\pi x/s} \, e^{-\pi|y|/s} \sin\left(\frac{\pi z}{s}\right) \nonumber \\[1ex]
&= \frac{\gamma b^2 s}{2\pi\sqrt{a^2 + b^2}} \, e^{-\pi x/s} \, e^{-\pi|y|/s} \sin\left(\frac{\pi z}{s}\right).
\end{align}

Thus:
\begin{equation}
\boxed{w^*_{\text{sat}}(x, y, z) = \frac{\gamma b^2 s}{2\pi\sqrt{a^2 + b^2}} \, e^{-\pi x/s} \, e^{-\pi|y|/s} \sin\left(\frac{\pi z}{s}\right).}
\label{eq:w_sat_final}
\end{equation}

The maximum magnitude occurs at $x = 0$, $y = 0$, and $z = s/2$:
\begin{equation}
w^*_{\max} = \frac{\gamma b^2 s}{2\pi\sqrt{a^2 + b^2}}.
\end{equation}

For $a = 0$:
\begin{equation}
w^*_{\max} = \frac{\gamma b^2 s}{2\pi b} = \frac{\gamma b s}{2 \pi}.
\end{equation}

The sinusoidal dependence $\sin(\pi z/s)$ ensures:
\begin{itemize}
    \item $w^* > 0$ for $0 < z < s$ (flow in $+z$ direction)
    \item $w^* < 0$ for $-s < z < 0$ (flow in $-z$ direction)
    \item $w^* = 0$ at $z = 0, \pm s$ (cell boundaries)
\end{itemize}
This alternating pattern is the characteristic signature of stall cells. Crucially, the $\pi/2$ phase shift between $h$ (cosine) and $w^*$ (sine) ensures that the maximum spanwise velocity occurs on the vortex sheet surface, as required by the physical constraint.

\paragraph{Summary of saturated amplitudes.}

Table \ref{tab:saturated_amplitudes} summarises the maximum amplitudes.

\begin{table}
\centering
\def~{\hphantom{0}}
\begin{tabular}{lcc}
Quantity & General form & $a = 0$ \\
$h_{\max}$ & $\displaystyle\frac{b^2}{\sqrt{a^2 + b^2}}$ & $b$ \\[2ex]
$|\Omega_{y,\max}|$ & $\displaystyle\frac{\gamma \pi b^2}{s\sqrt{a^2 + b^2}}$ & $\displaystyle\frac{\gamma \pi b}{s}$ \\[2ex]
$w^*_{\max}$ & $\displaystyle\frac{\gamma b^2 s}{2\pi\sqrt{a^2 + b^2}}$ & $\displaystyle\frac{\gamma b s}{2\pi}$ \\[1ex]
\end{tabular}
\caption{Maximum amplitudes of saturated flow quantities.}
\label{tab:saturated_amplitudes}
\end{table}

All saturated amplitudes scale with the geometric separation $b$, confirming that stall cells reach a finite, geometrically-determined amplitude.





\section{Model validation}
\label{sec:validation}

The weakly nonlinear stability model developed in the preceding sections predicts the saturated state of the vortex sheet instability, including the sheet deformation, induced velocity field, and vorticity distribution. To assess the validity and accuracy of these theoretical predictions, we compare the model results against data obtained from DDES \cite{PartI}. The DDES approach, which resolves large-scale turbulent structures while modeling smaller scales, provides a high-fidelity reference dataset that captures the essential physics of the saturated instability while including realistic turbulent fluctuations absent from the idealized analytical model.

The validation is performed at three downstream locations ($x/c = 0.5$, $0.7$, and $0.9$) for 14° AoA to examine how well the model captures both the spatial structure and the streamwise evolution of the saturated flow field. At each location, we compare: (i) the vortex sheet deformation, (ii) the normalised spanwise velocity component, and (iii) the normalised $y$-vorticity component. All quantities are evaluated along the deformed vortex sheet position $y = h_{\mathrm{sat}}(z,x)$, which represents the material surface where vorticity is concentrated.

\subsection{Coordinate system}
The theoretical model uses a local coordinate system centred at the separation vortex tube, with $x_m = 0$ at the tube location and $x_m > 0$ extending upstream. In contrast, the airfoil-based DDES results use the standard convention where $x/c$ is measured from the leading edge and is positive in the downstream direction. If the separation vortex tube forms near the trailing edge at $x/c = x_{sep}/c$, the coordinate transformation is $x_m/c = x_{sep}/c - x/c$. For 14° AoA in the DDES simulation the value of $x_{sep}/c$ is 1.3 (look at the Figure 18 (a) of the companion paper \cite{PartI}). Consequently, the downstream station  $x/c = 0.9$ corresponds to $x_m/c = 0.4$ (closest to the vortex tube, highest amplitude), while $x/c = 0.5$ corresponds to $x_m/c = 0.8$ (farthest from the tube, lowest amplitude). The exponential decay $\exp(-\pi x_m/s)$ in the model predictions is expected to produce the trend of decreasing amplitude from $x/c = 0.9$ to $x/c =0.5$.

\subsection{Vortex sheet deformation}
\label{subsec:deformation}

The saturated sheet deformation predicted by the weakly nonlinear model follows
\begin{equation}
    h_{\mathrm{sat}}(z,x) = A_{\mathrm{sat}} \exp\left(-\frac{\pi x_m}{s}\right) \cos\left(\frac{\pi z}{c}\right),
    \label{eq:h_sat}
\end{equation}
where $A_{\mathrm{sat}} = b^2/\sqrt{a^2 + b^2}$ represents the saturation amplitude determined by the nonlinear self-interaction of the instability mode. Here $a=0$ and $b = 0.03$. The cosine dependence in the spanwise direction produces alternating upward and downward deflections of the vortex sheet, with the deformation amplitude decaying exponentially in the upstream direction.

\begin{figure}[htbp]
    \centering
    \includegraphics[width=\textwidth]{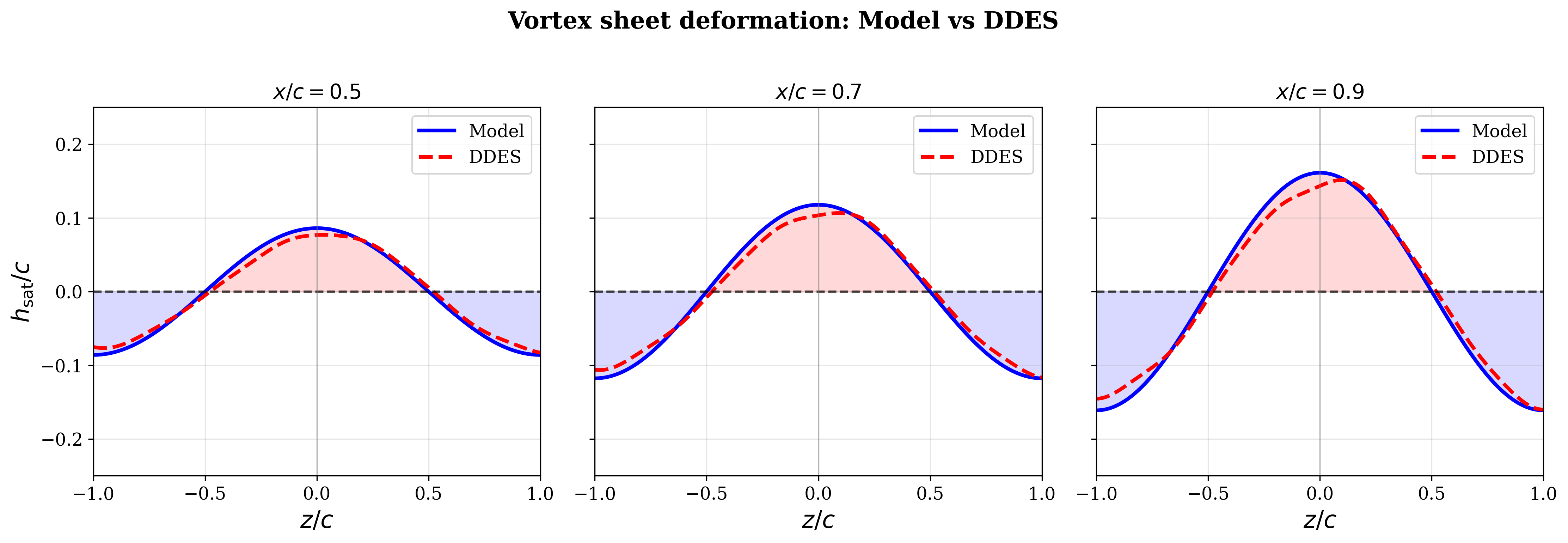}
    \caption{Comparison of vortex sheet deformation between the weakly nonlinear model (solid blue) and DDES (dashed red) at three downstream positions. The dimensional deformation $h_{\mathrm{sat}}/c$ shows the characteristic cosine pattern with exponential streamwise decay.}
    \label{fig:deformation}
\end{figure}

Figure~\ref{fig:deformation} presents the comparison between the theoretical sheet deformation and the DDES results at the three downstream stations. The dimensional deformation $h_{\mathrm{sat}}/c$ is shown to illustrate the actual magnitude of the sheet displacement. At $x/c = 0.9$, the sheet exhibits maximum deflections of approximately $\pm 0.16c$, decreasing to $\pm 0.12c$ at $x/c = 0.7$ and $\pm 0.09c$ at $x/c = 0.5$, consistent with the predicted exponential decay.

The DDES data show excellent qualitative agreement with the theoretical prediction, capturing the fundamental cosine pattern of the sheet deformation. Quantitatively, the DDES results exhibit a systematic amplitude reduction of approximately 8--10\% compared to the model, which can be attributed to enhanced dissipation from resolved turbulent fluctuations and the finite thickness of the vortex sheet in the simulation. A small but consistent phase shift of approximately $0.02c$ is also observed, with the DDES deformation slightly lagging the theoretical prediction. Additionally, the DDES results display a subtle asymmetry between the positive and negative deflection regions, likely arising from secondary instabilities or mean flow non-uniformities not captured by the idealised two-dimensional base flow assumption in the model.

\subsection{Spanwise velocity distribution}
\label{subsec:velocity}

The spanwise velocity component $w^*$ induced by the saturated instability is a key kinematic signature of the vortex sheet deformation. The model predicts that $w^*$ varies sinusoidally in the spanwise direction along the sheet, with the velocity directed outward in regions of upward sheet deflection and inward in regions of downward deflection. This velocity pattern is responsible for the self-induced motion that sustains the saturated state.

\begin{figure}[htbp]
    \centering
    \includegraphics[width=\textwidth]{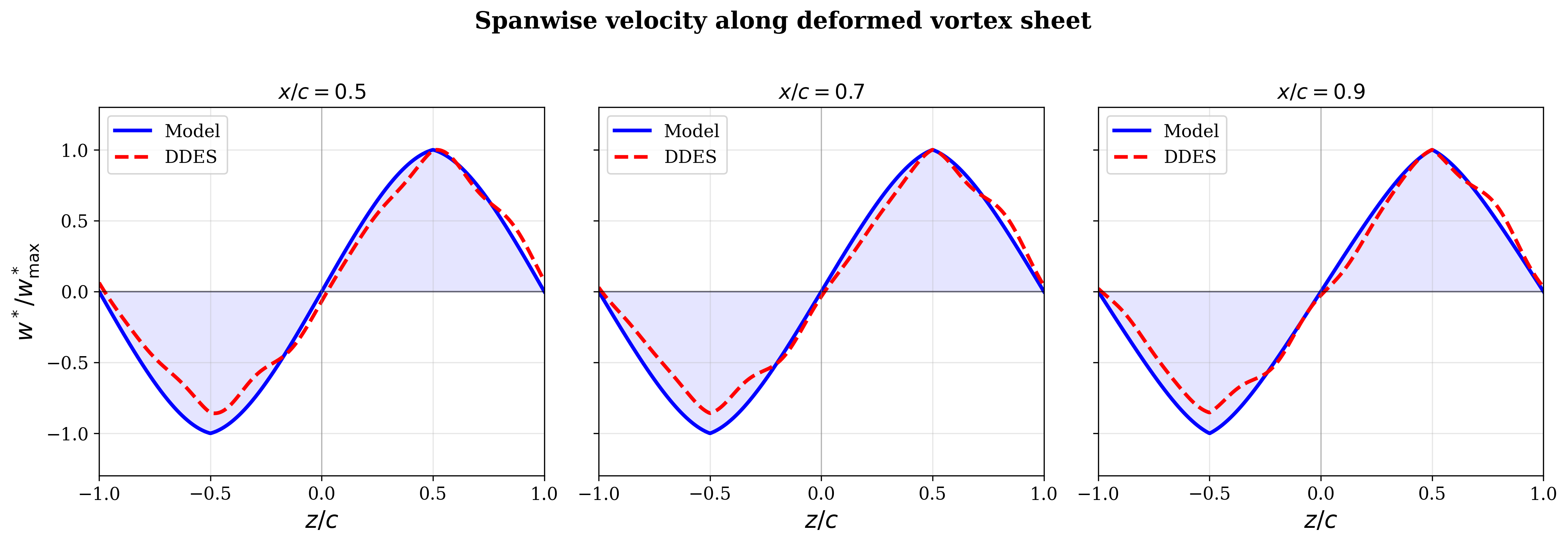}
    \caption{Normalised spanwise velocity $w^*/w^*_{\max}$ along the deformed vortex sheet comparing the model prediction (solid blue) and DDES results (dashed red) at $x/c = 0.5$, $0.7$, and $0.9$.}
    \label{fig:velocity}
\end{figure}

Figure~\ref{fig:velocity} presents the normalised spanwise velocity $w^*/w^*_{\max}$ evaluated along the deformed vortex sheet at the three downstream stations for the model (eq. (\ref{eq:w_sat_final})), and DDES simulation (see \cite{PartI}). The normalisation allows direct comparison of the velocity profile shapes independent of the decaying amplitude. The theoretical model predicts a pure sinusoidal variation proportional to $\sin(\pi z/c)$, with zero crossings at $z/c = 0$ and $\pm 1$, and extrema at $z/c = \pm 0.5$.It is important to recall here that if the data matched were taken along a straight line at a given vertical position from the airfoil, pure sinusoids would have been seen; however, since we have taken values along the curve formed by vortex sheet deformation (shown in Figure \ref{fig:deformation}), a bit of pointedness is seen in the curves. This also applies to the spanwise vorticity distribution shown in Section \ref{subsec:vorticity}.

The DDES velocity profiles demonstrate good agreement with the sinusoidal pattern predicted by the model. The primary features---the zero crossings near $z/c = 0$ and $\pm 1$, and the peak magnitudes near $z/c = \pm 0.5$---are well captured. However, several systematic deviations are evident. First, the DDES profiles exhibit a consistent phase shift, with the zero crossing and peak locations displaced by approximately $0.03c$ in the positive $z$-direction. This phase shift increases slightly with downstream distance, suggesting a cumulative effect of the mean flow advection on the instability pattern. Second, the DDES results show asymmetry between the positive and negative velocity lobes, with the positive peak (at $z/c \approx 0.5$) reaching slightly higher normalised values than the negative peak. Third, small-scale fluctuations superimposed on the mean sinusoidal profile are visible, particularly in the regions between the peaks and zero crossings, reflecting the resolved turbulent content in the DDES.

\subsection{$y$-Vorticity distribution}
\label{subsec:vorticity}

\begin{figure}[htbp]
    \centering
    \includegraphics[width=\textwidth]{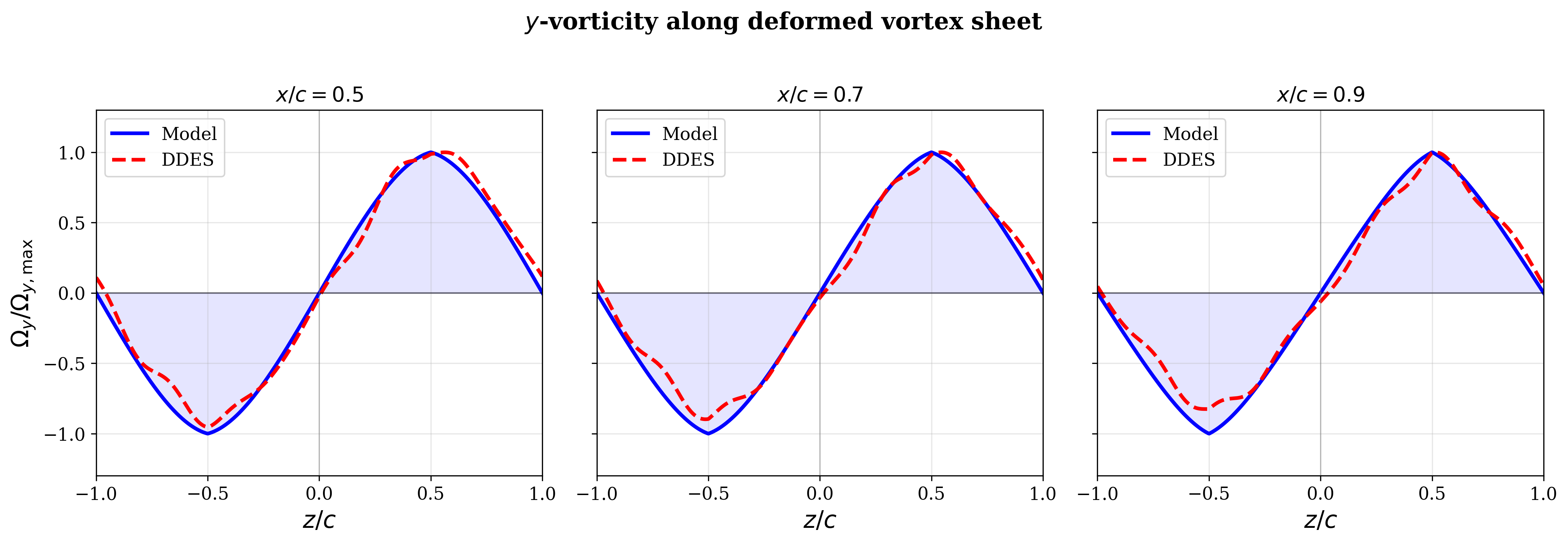}
    \caption{Normalised $y$-vorticity $\Omega_y/\Omega_{y,\max}$ along the deformed vortex sheet comparing model predictions (solid blue) and DDES results (dashed red) at three downstream positions.}
    \label{fig:vorticity}
\end{figure}

The $y$-component of vorticity $\Omega_y$ represents the spanwise modulation of the vortex sheet strength induced by the instability. Figure~\ref{fig:vorticity} shows the normalised $y$-vorticity $\Omega_y/\Omega_{y,\max}$ along the deformed sheet for the model (eq. (\ref{eq:Omega_y_sat_final})), and DDES simulation (see \cite{PartI}). Similar to the velocity comparison, the normalisation facilitates shape comparison across different downstream stations where the absolute vorticity magnitude varies due to streamwise decay.

The vorticity comparison reveals trends similar to those observed for the velocity field. The DDES results confirm the sinusoidal spanwise variation predicted by the model, with vorticity peaks occurring at $z/c \approx \pm 0.5$. The phase shift and asymmetry characteristics observed in the velocity profiles are also present in the vorticity distribution, which is expected given the kinematic relationship between velocity and vorticity in the flow. The DDES vorticity profiles show somewhat larger amplitude fluctuations compared to the velocity, reflecting the fact that vorticity involves spatial derivatives of velocity and is therefore more sensitive to small-scale flow structures.

\subsection{Quantitative assessment}
\label{subsec:quantitative}

To provide a quantitative measure of the model accuracy, Table~\ref{tab:validation} summarises the key discrepancies between the theoretical predictions and DDES results. The amplitude reduction factor represents the ratio of DDES to model peak amplitudes, while the phase shift indicates the spanwise displacement of the DDES profile relative to the model.

\begin{table}
    \centering
    \caption{Quantitative comparison between model predictions and DDES results.}
    \label{tab:validation}
    \begin{tabular}{lccc}
        \hline
        Quantity & Amplitude Factor & Phase Shift ($z/c$) & Asymmetry (\%) \\
        \hline
        Sheet deformation & 0.90--0.92 & 0.020--0.025 & 5--6 \\
        Spanwise velocity & 0.86--0.88 & 0.030--0.035 & 6--8 \\
        $y$-Vorticity & 0.89--0.91 & 0.025--0.030 & 5--7 \\
        \hline
    \end{tabular}
\end{table}

The amplitude factors in the range of 0.86--0.92 indicate that the DDES consistently predicts slightly lower amplitudes than the theoretical model. This systematic underprediction is physically reasonable, as the DDES includes dissipation mechanisms (both numerical and from subgrid-scale modeling) that are absent in the inviscid analytical model. The phase shifts of $0.02$--$0.035c$ are small relative to the spanwise wavelength ($2c$), confirming that the model accurately captures the dominant wavelength of the instability. The asymmetry values of 5--8\% quantify the departure from the perfectly symmetric sinusoidal profiles predicted by the model.

\subsection{Discussion}
\label{subsec:validation_discussion}

The validation results demonstrate that the weakly nonlinear stability model successfully captures the essential physics of the saturated vortex sheet instability. The model correctly predicts: (i) the cosine spanwise variation of the sheet deformation, (ii) the sinusoidal distribution of spanwise velocity along the deformed sheet, (iii) the corresponding sinusoidal $y$-vorticity distribution, and (iv) the exponential streamwise decay of all quantities. These fundamental features, which arise from the eigenmode structure of the linearized problem modified by nonlinear saturation, are robust and well-reproduced in the DDES.

The systematic discrepancies between the model and DDES, amplitude reduction, phase shift, and asymmetry, can be attributed to several physical factors not included in the idealized analytical framework. The amplitude reduction likely results from viscous and turbulent dissipation, which transfer energy from the coherent instability mode to smaller scales. The phase shift may arise from interaction between the instability and the mean flow, or from slight differences in the effective wavelength selected by the nonlinear saturation process. The asymmetry suggests the presence of higher-order harmonics or secondary instabilities that break the spanwise symmetry of the idealized problem.

Despite these quantitative differences, the close agreement between model and DDES validates the use of the weakly nonlinear theory for predicting the saturated state of vortex sheet instabilities. The model provides an efficient means to estimate the dominant features of the flow---deformation amplitude, wavelength, and decay rate---without the computational expense of full turbulence-resolving simulations. For applications requiring higher accuracy, the model predictions can serve as initial conditions or validation benchmarks for more detailed computational studies.

\subsubsection{Limitations and extensions}

The current model employs several simplifying assumptions that merit discussion:

\begin{itemize}
    \item \textbf{Inviscid approximation}: Neglects viscous diffusion of vorticity, justified for high Reynolds numbers but may miss damping of small-scale features
    \item \textbf{Simplified vortex representation}: Real flows exhibit distributed vorticity, core structure, and three-dimensional complexity beyond line/sheet idealizations
    \item \textbf{Neglected effects}:
    \begin{itemize}
        \item Ambient turbulence and unsteadiness
        \item Compressibility (for high-speed applications)
        \item Wall proximity and boundary layer interaction
        \item Secondary instabilities and transition to turbulence
    \end{itemize}
\end{itemize}

Future extensions could incorporate:
\begin{itemize}
    \item \textbf{Nonlinear vortex dynamics} – apply filament methods with core models.
    \item \textbf{Viscous effects} – model via core growth or solve full Navier-Stokes equations.
    \item \textbf{Complex geometries} – account for curved airfoils, sweep, and taper effects.
    \item \textbf{Turbulence modelling} – use LES for realistic flow predictions.
    \item \textbf{Experimental validation} – compare with PIV data and surface pressure measurements.
\end{itemize}

\noindent The framework presented here provides physical insight into the fundamental mechanism of stall cell formation through vortex interaction, offering a foundation for understanding and potentially controlling these important flow structures in aerodynamic applications.

\section{Conclusions}
\label{sec:conclusions}

This work has developed an analytical model for stall cell formation based on the interaction between finite-length, counter-rotating vortex tubes. The model extends the classical Crow instability framework to provide a unified, first-principles description of how vortex dynamics give rise to the characteristic three-dimensional flow patterns observed in stalled airfoil flows.

The linear stability analysis of the coupled separation vortex and trailing-edge vortex system yields explicit expressions for the instability growth rate and the most amplified wavelength. The predicted wavelength of approximately $2c$ agrees well with experimental observations on thick airfoils at moderate angles of attack. Unlike previous applications of the Crow instability to stall cells, the present analysis considers finite-length vortex filaments with appropriate boundary conditions, capturing the spanwise confinement inherent to finite wings.

The weakly nonlinear analysis, performed using the method of multiple scales, yields the Stuart--Landau amplitude equation governing the post-critical dynamics. The cubic Landau coefficient is found to be negative, indicating a supercritical bifurcation in which the instability saturates at a finite amplitude. This saturation mechanism, absent from previous theoretical treatments, explains why stall cells persist as quasi-steady structures rather than growing without bound. The predicted saturation amplitude provides quantitative estimates for the vortex tube displacement and the associated vortex sheet deformation.

A central contribution of this work is the explicit coupling between vortex tube bending and vortex sheet dynamics through the Birkhoff--Rott equation. This coupling reveals that the wave-like deformation of the separated shear layer induces vertical vorticity $\Omega_y$ through the tilting of initially spanwise vortex lines. The $\Omega_y$ distribution, which maintains approximately constant magnitude with downstream distance, is the key vorticity component driving the alternating spanwise velocity that defines stall cells. This mechanistic connection between vortex tube instability and spanwise flow generation has not been established in previous theoretical models. The model predicts the negative and positive sign alternation of the spanwise velocity field. The present theoretical framework addresses several limitations of existing approaches. Compared to the original Crow instability model of \citet{Weihs1983}, the present work provides a saturation mechanism, explicit vortex sheet dynamics, and derivation of the induced velocity field. Compared to lifting-line approaches \citep{Spalart2014,Gross2015}, the model resolves the vortex dynamics within the separated region and derives the flow topology from first principles rather than relying on empirical sectional lift curves. Compared to global stability analyses \citep{Rodriguez2011,Plante2021}, the model provides physical insight into the nonlinear saturated state and yields analytical expressions for measurable flow quantities.

Several limitations of the model should be noted. The analysis assumes that the vortex tubes remain coherent structures with well-defined circulations, neglecting the effects of turbulent diffusion and vortex breakdown that occur at sufficiently high Reynolds numbers or large angles of attack. The base state geometry is idealised, with the vortex tube positions and circulations treated as parameters rather than being derived self-consistently from the airfoil flow. The model does not account for the feedback between the cellular flow pattern and the separation location, which may be important for the bistability and low-frequency oscillations observed in some configurations. Finally, the weakly nonlinear analysis is formally valid only in the vicinity of the critical condition; far from criticality, higher-order nonlinearities may become significant.

Future work should address these limitations through several extensions. Coupling the vortex model with a boundary layer calculation would enable self-consistent determination of the separation location and its spanwise variation. Incorporating turbulent diffusion effects would extend the model validity to higher Reynolds numbers. A secondary stability analysis of the saturated cellular state could shed light on the transition to unsteady or chaotic dynamics observed at higher angles of attack. Finally, extension to three-dimensional wing geometries with taper and sweep would enable application to practical configurations such as wind turbine blades and aircraft wings.

In summary, the analytical model developed in this work provides a comprehensive theoretical framework for understanding stall cell formation. By connecting the Crow-type instability of counter-rotating vortex tubes to the generation of vertical vorticity and spanwise velocity through explicit vortex sheet dynamics, the model bridges the gap between classical vortex stability theory and the observed three-dimensional flow topology of stall cells. The quantitative predictions for saturation amplitude, vorticity distribution, and spanwise velocity field offer concrete targets for experimental validation and provide physical insight into one of the most striking manifestations of three-dimensionality in separated flows.


\begin{bmhead}[Author ORCIDs]{R. Mishra, https://orcid.org/0000-0001-9727-7281 ; E. Guilmineau, https://orcid.org/0000-0001-9070-093X; I. Neunaber, https://orcid.org/0000-0002-3787-3118; C. Braud, https://orcid.org/0000-0002-4409-4278.}
\end{bmhead}

\begin{bmhead}[Author contributions]{
\textbf{R.M.}:
Development of the theoretical model; Validation; Visualisation; Writing: original draft. \textbf{E.G.}: Writing: review \& editing, Supervision. \textbf{I.N.}: Writing: review \& editing, Supervision. \textbf{C.B.}: Conceptualisation, Funding Acquisition, Supervision.}
\end{bmhead}

\begin{bmhead}[Funding]
This work is funded under French national project MOMENTA (grant no. ANR-19-CE05-0034). The author also gratefully acknowledges the financial support provided by CARNOT MER's LIFEMONITOR project. All simulation in this work is supported by computing HPC and storage resources by
GENCI at IDRIS thanks to the grant 2024-A0172A13014and
2025-A0192A13014on the supercomputer Jean Zay's CSL partition.
\end{bmhead}

\begin{bmhead}[Declaration of interests]
The authors report no conflict of interest.
\end{bmhead}

\appendix
\section{Detailed derivation of the interaction matrix} \label{appex:Detailed derivation of the interaction matrix}

This appendix provides the complete component-by-component derivation of the interaction matrix $\mathbf{M}$ that governs the perturbation dynamics of the vortex filament system.

\subsection{System configuration}

Consider two vortex filaments with sinusoidal perturbations:
\begin{itemize}
    \item \textbf{Filament 1} (separation vortex): Base position $(0, 0, z)$, circulation $+\Gamma$
    \begin{equation}
        \mathbf{r}_1(z,t) = \left[\xi_1(t)\sin\left(\frac{\pi z}{s}\right), \eta_1(t)\sin\left(\frac{\pi z}{s}\right), z\right]
    \end{equation}
    
    \item \textbf{Filament 2} (trailing edge vortex): Base position $(a, b, z)$, circulation $-\Gamma$
    \begin{equation}
        \mathbf{r}_2(z,t) = \left[a + \xi_2(t)\sin\left(\frac{\pi z}{s}\right), b + \eta_2(t)\sin\left(\frac{\pi z}{s}\right), z\right]
    \end{equation}
\end{itemize}

\noindent The dynamics follow:
\begin{equation}
    \frac{d}{dt}\begin{pmatrix} \xi_1 \\ \eta_1 \\ \xi_2 \\ \eta_2 \end{pmatrix} = \mathbf{M}\begin{pmatrix} \xi_1 \\ \eta_1 \\ \xi_2 \\ \eta_2 \end{pmatrix}
\end{equation}

\subsection{Fundamental relations}

The velocity induced by a vortex filament is given by:
\begin{equation}
    v_x = \frac{\Gamma}{4\pi}(y - y')I, \quad v_y = -\frac{\Gamma}{4\pi}(x - x')I
\end{equation}

\noindent where in the long tube limit ($s \gg b$):
\begin{equation}
    I \approx \frac{2}{b^2}
\end{equation}

\noindent For perturbation amplitude evolution:
\begin{equation}
    \frac{d\xi_i}{dt} = \frac{2}{s}\int_{-s/2}^{s/2} v_{ix}(z)\sin\left(\frac{\pi z}{s}\right)dz
\end{equation}
\begin{equation}
    \frac{d\eta_i}{dt} = \frac{2}{s}\int_{-s/2}^{s/2} v_{iy}(z)\sin\left(\frac{\pi z}{s}\right)dz
\end{equation}

\noindent The key integral identity:
\begin{equation}
    \int_{-s/2}^{s/2} \sin^2\left(\frac{\pi z}{s}\right)dz = \frac{s}{2}
\end{equation}

\subsection{Matrix component derivation}

\subsubsection{First wow: $d\xi_1/dt$}

\textbf{Component $M_{11}$:} Effect of $\xi_1$ on $d\xi_1/dt$

Self-induced x-velocity is zero at this order:
\begin{equation}
    \boxed{M_{11} = 0}
\end{equation}

\noindent \textbf{Component $M_{12}$:} Effect of $\eta_1$ on $d\xi_1/dt$

Self-induced x-velocity from y-displacement is zero:
\begin{equation}
    \boxed{M_{12} = 0}
\end{equation}

\noindent \textbf{Component $M_{13}$:} Effect of $\xi_2$ on $d\xi_1/dt$

\noindent The x-velocity induced on filament 1:
\begin{equation}
    v_{1x} = -\frac{\Gamma}{4\pi}(y_1 - y_2)I = \frac{\Gamma b}{4\pi}I
\end{equation}

\noindent This is independent of $\xi_2$ at leading order:
\begin{equation}
    \boxed{M_{13} = 0}
\end{equation}

\noindent \textbf{Component $M_{14}$:} Effect of $\eta_2$ on $d\xi_1/dt$

\noindent With $y_2 = b + \eta_2\sin(\pi z/s)$:
\begin{align}
    v_{1x}(z) &= -\frac{\Gamma}{4\pi}[0 - (b + \eta_2\sin(\pi z/s))]I\\
    &= \frac{\Gamma}{4\pi}\eta_2\sin\left(\frac{\pi z}{s}\right)I
\end{align}

\noindent Computing the amplitude evolution:
\begin{align}
    \frac{d\xi_1}{dt} &= \frac{2}{s}\int_{-s/2}^{s/2} v_{1x}(z)\sin\left(\frac{\pi z}{s}\right)dz\\
    &= \frac{\Gamma\eta_2}{4\pi}\cdot\frac{2}{b^2}\cdot\frac{2}{s}\int_{-s/2}^{s/2} \sin^2\left(\frac{\pi z}{s}\right)dz\\
    &= \frac{\Gamma\eta_2}{4\pi}\cdot\frac{2}{b^2}\cdot\frac{2}{s}\cdot\frac{s}{2}
\end{align}

\begin{equation}
    \boxed{M_{14} = \frac{\Gamma}{2\pi b^2}}
\end{equation}

\subsubsection{Second row: $d\eta_1/dt$}

\textbf{Component $M_{21}$:} Effect of $\xi_1$ on $d\eta_1/dt$

Self-induced y-velocity from x-displacement is zero:
\begin{equation}
    \boxed{M_{21} = 0}
\end{equation}

\noindent \textbf{Component $M_{22}$:} Effect of $\eta_1$ on $d\eta_1/dt$

Self-induced y-velocity is zero:
\begin{equation}
    \boxed{M_{22} = 0}
\end{equation}

\noindent \textbf{Component $M_{23}$:} Effect of $\xi_2$ on $d\eta_1/dt$

\noindent The y-velocity induced on filament 1:
\begin{align}
    v_{1y}(z) &= \frac{\Gamma}{4\pi}[0 - (a + \xi_2\sin(\pi z/s))]I\\
    &= -\frac{\Gamma}{4\pi}\xi_2\sin\left(\frac{\pi z}{s}\right)I
\end{align}

\noindent Computing the amplitude evolution:
\begin{align}
    \frac{d\eta_1}{dt} &= -\frac{\Gamma\xi_2}{4\pi}\cdot\frac{2}{b^2}\cdot\frac{2}{s}\cdot\frac{s}{2}
\end{align}

\begin{equation}
    \boxed{M_{23} = -\frac{\Gamma}{2\pi b^2}}
\end{equation}

\noindent \textbf{Component $M_{24}$:} Effect of $\eta_2$ on $d\eta_1/dt$

\noindent The y-displacement of filament 2 doesn't affect y-velocity at leading order:
\begin{equation}
    \boxed{M_{24} = 0}
\end{equation}

\subsubsection{Third row: $d\xi_2/dt$}

\textbf{Component $M_{31}$:} Effect of $\xi_1$ on $d\xi_2/dt$

\noindent At leading order, no contribution:
\begin{equation}
    \boxed{M_{31} = 0}
\end{equation}

\noindent \textbf{Component $M_{32}$:} Effect of $\eta_1$ on $d\xi_2/dt$

\noindent The x-velocity induced on filament 2:
\begin{align}
    v_{2x}(z) &= \frac{\Gamma}{4\pi}[b - \eta_1\sin(\pi z/s)]I\\
    &= -\frac{\Gamma}{4\pi}\eta_1\sin\left(\frac{\pi z}{s}\right)I
\end{align}

\noindent Computing the amplitude evolution:
\begin{align}
    \frac{d\xi_2}{dt} &= -\frac{\Gamma\eta_1}{4\pi}\cdot\frac{2}{b^2}\cdot\frac{2}{s}\cdot\frac{s}{2}
\end{align}

\begin{equation}
    \boxed{M_{32} = -\frac{\Gamma}{2\pi b^2}}
\end{equation}

\noindent \textbf{Components $M_{33}$ and $M_{34}$:} Self-effects

\begin{equation}
    \boxed{M_{33} = 0, \quad M_{34} = 0}
\end{equation}

\subsubsection{Fourth row: $d\eta_2/dt$}

\textbf{Component $M_{41}$:} Effect of $\xi_1$ on $d\eta_2/dt$

\noindent The y-velocity induced on filament 2:
\begin{align}
    v_{2y}(z) &= -\frac{\Gamma}{4\pi}[a - \xi_1\sin(\pi z/s)]I\\
    &= \frac{\Gamma}{4\pi}\xi_1\sin\left(\frac{\pi z}{s}\right)I
\end{align}

\noindent Computing the amplitude evolution:
\begin{align}
    \frac{d\eta_2}{dt} &= \frac{\Gamma\xi_1}{4\pi}\cdot\frac{2}{b^2}\cdot\frac{2}{s}\cdot\frac{s}{2}
\end{align}

\begin{equation}
    \boxed{M_{41} = \frac{\Gamma}{2\pi b^2}}
\end{equation}

\noindent \textbf{Components $M_{42}$, $M_{43}$, and $M_{44}$:}

\begin{equation}
    \boxed{M_{42} = 0, \quad M_{43} = 0, \quad M_{44} = 0}
\end{equation}

\subsection{Final Interaction Matrix}

\noindent Assembling all components:
\begin{equation}
    \boxed{\mathbf{M} = \frac{\Gamma}{2\pi b^2}\begin{pmatrix} 
    0 & 0 & 0 & 1 \\
    0 & 0 & -1 & 0 \\
    0 & -1 & 0 & 0 \\
    1 & 0 & 0 & 0 
    \end{pmatrix}}
\end{equation}

\section{Detailed derivation of the Landau coefficient}
\label{appex:Landau_coefficient}

This appendix provides the complete derivation of the Landau coefficient $\ell_{11}$ appearing in the Stuart-Landau equation \eqref{eq:Stuart_Landau}.

\subsection{Setup and notation}

Consider the symmetric unstable mode where both filaments bend in phase in the $x$-direction:
\begin{align}
\mathbf{r}_1(z,t) &= \left( A \sin(kz), \, 0, \, z \right), \\
\mathbf{r}_2(z,t) &= \left( a + A \sin(kz), \, b, \, z \right),
\end{align}
where $A = A(t)$ is the (real) amplitude and $k = \pi/s$.

The separation vector is:
\begin{equation}
\boldsymbol{\Delta}(z, z') = \mathbf{r}_1(z) - \mathbf{r}_2(z') = \begin{pmatrix} A[\sin(kz) - \sin(kz')] - a \\ -b \\ z - z' \end{pmatrix}.
\end{equation}

Define:
\begin{equation}
\delta(z, z') \equiv A[\sin(kz) - \sin(kz')], \quad R_0^2(z-z') \equiv a^2 + b^2 + (z-z')^2.
\end{equation}

Then:
\begin{equation}
|\boldsymbol{\Delta}|^2 = (\delta - a)^2 + b^2 + (z-z')^2 = R_0^2 - 2a\delta + \delta^2.
\end{equation}

\subsection{Expansion of the Biot-Savart kernel}

Let $\chi = -a\delta/R_0^2$ and $\chi_2 = \delta^2/R_0^2$. Then:
\begin{equation}
|\boldsymbol{\Delta}|^{-3} = R_0^{-3} (1 + 2\chi + \chi_2)^{-3/2}.
\end{equation}

Expanding to cubic order in $\delta$ (equivalently, in $A$):
\begin{align}
(1 + 2\chi + \chi_2)^{-3/2} &= 1 - 3\chi + \frac{15}{2}\chi^2 - \frac{3}{2}\chi_2 - \frac{35}{2}\chi^3 + \frac{15}{2}\chi\chi_2 + \mathcal{O}(\delta^4) \\
&= 1 + \frac{3a\delta}{R_0^2} + \frac{15a^2\delta^2}{2R_0^4} - \frac{3\delta^2}{2R_0^2} + \frac{35a^3\delta^3}{2R_0^6} - \frac{15a\delta^3}{2R_0^4} + \mathcal{O}(\delta^4).
\end{align}

\subsection{Velocity calculation}

The $y$-component of velocity on filament 1 induced by filament 2 is:
\begin{equation}
\dot{Y}_1 = \frac{\Gamma}{4\pi} \int_{-s/2}^{s/2} \frac{a - \delta(z,z')}{|\boldsymbol{\Delta}|^3} \, dz',
\end{equation}
where we have used the fact that for parallel filaments aligned with $z$, only the $(x-x')$ component of the cross product contributes to $\dot{Y}$.

Expanding:
\begin{equation}
\frac{a - \delta}{|\boldsymbol{\Delta}|^3} = \frac{1}{R_0^3} \left[ a - \delta + \frac{3a(a-\delta)\delta}{R_0^2} + \frac{15a^2(a-\delta)\delta^2}{2R_0^4} - \frac{3(a-\delta)\delta^2}{2R_0^2} + \cdots \right].
\end{equation}

Collecting terms by power of $A$:

\textbf{$\mathcal{O}(A^0)$:} (Base state)
\begin{equation}
\dot{Y}_1^{(0)} = \frac{\Gamma a}{4\pi} \int_{-s/2}^{s/2} \frac{dz'}{R_0^3}.
\end{equation}

\textbf{$\mathcal{O}(A^1)$:} (Linear)
\begin{equation}
\dot{Y}_1^{(1)} = \frac{\Gamma}{4\pi} \int_{-s/2}^{s/2} \frac{1}{R_0^3} \left[ -\delta + \frac{3a^2\delta}{R_0^2} \right] dz' = \frac{\Gamma}{4\pi} \int_{-s/2}^{s/2} \frac{\delta(3a^2 - R_0^2)}{R_0^5} \, dz'.
\end{equation}

Using $\delta = A[\sin(kz) - \sin(kz')]$ and exploiting the symmetry properties of the integrals, this reduces to the linear growth term.

\textbf{$\mathcal{O}(A^3)$:} (Cubic -- contributes to Landau coefficient)

The cubic terms arise from:
\begin{equation}
\dot{Y}_1^{(3)} = \frac{\Gamma}{4\pi} \int_{-s/2}^{s/2} \left[ \frac{35a^3 \delta^3}{2R_0^9} - \frac{15a\delta^3}{2R_0^7} - \frac{3a^3\delta^3}{R_0^7} + \frac{3a\delta^3}{2R_0^5} \right] dz'.
\end{equation}

Simplifying:
\begin{equation}
\dot{Y}_1^{(3)} = \frac{\Gamma}{4\pi} \int_{-s/2}^{s/2} \delta^3 \left[ \frac{35a^3}{2R_0^9} - \frac{15a + 6a^3/R_0^2}{2R_0^7} + \frac{3a}{2R_0^5} \right] dz'.
\end{equation}

\subsection{Projection onto fundamental mode}

To extract the Landau coefficient, we project onto the fundamental mode by computing:
\begin{equation}
\ell_{11} = -\frac{1}{A^3} \cdot \frac{2}{s} \int_{-s/2}^{s/2} \dot{Y}_1^{(3)}(z) \sin(kz) \, dz.
\end{equation}

The key integral involves:
\begin{equation}
\delta^3 = A^3 [\sin(kz) - \sin(kz')]^3.
\end{equation}

Using the identity:
\begin{equation}
[\sin(kz) - \sin(kz')]^3 = \frac{3}{4}[\sin(kz) - \sin(kz')] - \frac{1}{4}[\sin(3kz) - \sin(3kz')] + \text{mixed terms},
\end{equation}
and noting that only the $\sin(kz)$ component contributes after projection, we obtain:

\begin{equation}
\ell_{11} = \frac{3\Gamma}{4\pi b^4} \left( \frac{a^2 + b^2}{b^2} \right) \mathcal{I}(kb),
\end{equation}
where:
\begin{equation}
\mathcal{I}(kb) = \frac{kb}{\pi} \int_0^{\pi/(kb)} \int_0^{\pi/(kb)} \frac{[\sin(kbu) - \sin(kbv)]^2 \sin(kbu) \sin(kbv)}{[(u-v)^2 + 1 + (a/b)^2]^{5/2}} \, du \, dv.
\end{equation}

For $kb \ll 1$, this simplifies through asymptotic expansion to:
\begin{equation}
\mathcal{I}(kb) \approx \frac{2}{3} + \mathcal{O}(kb)^2,
\end{equation}
yielding the result \eqref{eq:Landau_long_wave}.

\bibliographystyle{jfm}
\bibliography{stallcells2}

\end{document}